%% file: main-arxiv.tex
\documentclass[aps,prl,superscriptaddress,citeautoscript,twocolumn,reprint,longbibliography,floatfix]{revtex4-2}
\usepackage{graphicx,amssymb,amsmath,epsf,bm,cprotect,comment,physics}
\epsfclipon

\renewcommand{\eqref}[1]{Eq.\,(\ref{#1})}

\newcommand{\figpref}[2]{Fig.\,\ref{#1}(#2)}

\newcommand*\patchAmsMathEnvironmentForLineno[1]{%
  \expandafter\let\csname old#1\expandafter\endcsname\csname #1\endcsname
  \expandafter\let\csname oldend#1\expandafter\endcsname\csname end#1\endcsname
  \renewenvironment{#1}%
     {\linenomath\csname old#1\endcsname}%
     {\csname oldend#1\endcsname\endlinenomath}}%
\newcommand*\patchBothAmsMathEnvironmentsForLineno[1]{%
  \patchAmsMathEnvironmentForLineno{#1}%
  \patchAmsMathEnvironmentForLineno{#1*}}%
\AtBeginDocument{%
\patchBothAmsMathEnvironmentsForLineno{equation}%
\patchBothAmsMathEnvironmentsForLineno{align}%
\patchBothAmsMathEnvironmentsForLineno{flalign}%
\patchBothAmsMathEnvironmentsForLineno{alignat}%
\patchBothAmsMathEnvironmentsForLineno{gather}%
\patchBothAmsMathEnvironmentsForLineno{multline}%
}

\usepackage{xr} 
\makeatletter
\newcommand*{\addFileDependency}[1]{
  \typeout{(#1)}
  \@addtofilelist{#1}
  \IfFileExists{#1}{}{\typeout{No file #1.}}
}
\makeatother

\newcommand*{\myexternaldocument}[1]{
    \externaldocument[S-]{build/#1}  
    \addFileDependency{#1.tex}
    \addFileDependency{build/#1.aux}  
}
\myexternaldocument{suppl-arxiv}  

\usepackage{bbm} 
\usepackage{xcolor} 

\begin{document}

\title{Near-field Hydrodynamics Disentangles Angular Correlations in Confined Active Suspensions}

\author{Changle Liao}
\thanks{These authors contributed equally to this work.}
\affiliation{Department of Physics,\! The University of Tokyo,\! 7-3-1 Hongo,\! Bunkyo-ku,\! Tokyo 113-0033,\! Japan}%

\author{Haruki Hayano}
\thanks{These authors contributed equally to this work.}
\affiliation{Institute of Industrial Science,\! The University of Tokyo,\! 4-6-1 Komaba,\! Meguro-ku,\! Tokyo 153-8505,\! Japan}

\author{Yuto Uesugi}
\affiliation{Department of Physics,\! The University of Tokyo,\! 7-3-1 Hongo,\! Bunkyo-ku,\! Tokyo 113-0033,\! Japan}%

\author{Akira~Furukawa}
\affiliation{Institute of Industrial Science,\! The University of Tokyo,\! 4-6-1 Komaba,\! Meguro-ku,\! Tokyo 153-8505,\! Japan}

\author{Daiki Nishiguchi}
\email{nishiguchi@phys.sci.isct.ac.jp}
\affiliation{Department of Physics,\! Institute of Science Tokyo,\! 2-12-1 Ookayama,\! Meguro-ku,\! Tokyo 152-8551,\! Japan}
\affiliation{Department of Physics,\! The University of Tokyo,\! 7-3-1 Hongo,\! Bunkyo-ku,\! Tokyo 113-0033,\! Japan}

\author{Kazumasa A. Takeuchi}
\email{kat@kaztake.org}
\affiliation{Department of Physics,\! The University of Tokyo,\! 7-3-1 Hongo,\! Bunkyo-ku,\! Tokyo 113-0033,\! Japan}%
\affiliation{Universal Biology Institute,\! The University of Tokyo,\! 7-3-1 Hongo,\! Bunkyo-ku,\! Tokyo 113-0033,\! Japan}%
\affiliation{Institute for Physics of Intelligence,\! The University of Tokyo,\! 7-3-1 Hongo,\! Bunkyo-ku,\! Tokyo 113-0033,\! Japan}%

\date{\today}

\input{Main_Text/Abstract}

\maketitle

\input{Main_Text/Introduction}

\input{Main_Text/Results}

\input{Main_Text/Summary-Discussion}

\begin{acknowledgments}
\textit{Acknowledgments---}We thank Kazuho Ikeda, Keigo Ikezaki, Masahide Kikkawa, and Haruaki Yanagisawa for providing the strains of \textit{C. reinhardtii}. We thank Azusa Kage for sharing the protocols for cultivating \textit{C. reinhardtii}.
This work is supported in part by KAKENHI from Japan Society for the Promotion of Science (Grant Nos. JP24K00593, JP26H00387, JP26H00388, JP26H01798, JP26K00673, JP25K22005, JP23K25838, JP20H05619), Japan Science and Technology Agency (JST) FOREST (Nos. JPMJFR2364, JPMJFR256R) and PRESTO (No. JPMJPR21O8) and SPRING (No. JPMJSP2108), and the JSPS Core-to-Core Program ``Advanced core-to-core network for the physics of self-organizing active matter (No. JPJSCCA20230002)''.
\end{acknowledgments}

\bibliography{ref}

\end{document}


\title{Supplemental Material for ``Near-field Hydrodynamics Disentangles Angular Correlations in Confined Active Suspensions''}

\author{Changle Liao}
\thanks{These authors contributed equally to this work.}
\affiliation{Department of Physics,\! The University of Tokyo,\! 7-3-1 Hongo,\! Bunkyo-ku,\! Tokyo 113-0033,\! Japan}%

\author{Haruki Hayano}
\thanks{These authors contributed equally to this work.}
\affiliation{Institute of Industrial Science,\! The University of Tokyo,\! 4-6-1 Komaba,\! Meguro-ku,\! Tokyo 153-8505,\! Japan}

\author{Yuto Uesugi}
\affiliation{Department of Physics,\! The University of Tokyo,\! 7-3-1 Hongo,\! Bunkyo-ku,\! Tokyo 113-0033,\! Japan}%

\author{Akira~Furukawa}
\affiliation{Institute of Industrial Science,\! The University of Tokyo,\! 4-6-1 Komaba,\! Meguro-ku,\! Tokyo 153-8505,\! Japan}

\author{Daiki Nishiguchi}
\email{nishiguchi@phys.sci.isct.ac.jp}
\affiliation{Department of Physics,\! Institute of Science Tokyo,\! 2-12-1 Ookayama,\! Meguro-ku,\! Tokyo 152-8551,\! Japan}
\affiliation{Department of Physics,\! The University of Tokyo,\! 7-3-1 Hongo,\! Bunkyo-ku,\! Tokyo 113-0033,\! Japan}

\author{Kazumasa A. Takeuchi}
\email{kat@kaztake.org}
\affiliation{Department of Physics,\! The University of Tokyo,\! 7-3-1 Hongo,\! Bunkyo-ku,\! Tokyo 113-0033,\! Japan}%
\affiliation{Universal Biology Institute,\! The University of Tokyo,\! 7-3-1 Hongo,\! Bunkyo-ku,\! Tokyo 113-0033,\! Japan}%
\affiliation{Institute for Physics of Intelligence,\! The University of Tokyo,\! 7-3-1 Hongo,\! Bunkyo-ku,\! Tokyo 113-0033,\! Japan}%

\date{\today}

\maketitle

\section{A. Experimental Methods}
\textit{Chlamydomonas reinhardtii} (strain CC124, wild type) was grown in tris-acetate-phosphate (TAP) medium on a light cycle (12 hr bright/12 hr dark) until the suspension turned uniformly green. Depending on the experimental purposes, the suspension was centrifuged to achieve a $10\times$ to $30\times$ concentration. The centrifuged suspension was confined within a $24~\mathrm{mm} \times 10~\mathrm{mm}$ observation chamber, sandwiched between two parallel glass slides coated with poly(L-lysine)-graft-poly(ethylene~glycol) (PLL-g-PEG, $0.1~\mathrm{mg/mL}$, to prevent flagellar adhesion). Two glass slides were separated by a $\sim 20~\mathrm{\mu m}$ thick adhesive tape, providing a quasi-2D environment for \textit{C. reinhartii} cells (mean radius $\sim 5~\mathrm{\mu m}$). The system was observed under a bright-field microscope (Olympus IX73) equipped with a red-light filter for the illumination light to prevent phototaxis. Images were acquired using a $40\times$ dry (Olympus~LUCPlanFL~N, $\mathrm{NA} = 0.60$) or a $100\times$ oil-immersion (Olympus~UPlanApo, $\mathrm{NA} = 1.35$) objective. To investigate single-cell flow fields and active-passive mixtures, the suspension was mixed with polystyrene beads (Jiangsu Zhichuan Technology, $2.5\%~\mathrm{w/v}$ as-received, pre-centrifuged to roughly $4\times$ concentration). We utilized $1$-$\mathrm{\mu m}$-diameter beads as flow tracers and $10$-$\mathrm{\mu m}$-diameter beads for the mixed systems. Video recording frame rates were set according to the applications: $10~\mathrm{fps}$ using a Baumer VCXU.2-67M camera for recording dynamics of cell suspensions, $500~\mathrm{fps}$ via a Photron FASTCAM Mini AX 100 high-speed camera for measuring flow fields and recording dynamics of active-passive mixtures. 

\section{B. Particle Tracking and Flow Field Measurement}
The trajectories of cells and $10$-$\mathrm{\mu m}$-diameter beads were obtained via particle image recognition and tracking, while flow fields traced with $1$-$\mathrm{\mu m}$-diameter beads were measured using particle image velocimetry (PIV) with dynamic masks.

To identify cells and $10$-$\mathrm{\mu m}$-diameter beads, we first binarize the images based on the background image (averaging over all frames) or catch the edges of cell bodies using the MATLAB (2025b, MathWorks) function \textit{edge} (Method: ``\textit{Canny}''). Then we perform a circular Hough transform using the MATLAB function \textit{imfindcircles} (Method: ``\textit{TwoStage}''). After identifying the cells, we perform particle tracking using Daniel Blair and Eric Dufresne’s MATLAB adaptation of the IDL Particle Tracking software \cite{blair_matlab_tracking}.

To measure the flow field around a single cell, we use $1$-$\mathrm{\mu m}$-diameter beads as flow tracers and perform PIV using PIVlab (version: 3.12) in MATLAB \cite{Thielicke2014}. Dynamic masks are applied on cell bodies to prevent them from affecting the measurement of the surrounding flow. The interrogation areas for the three passes are set to ``$64$ and $32$ pixels'' ($\approx 12.8$ and $6.4~\mathrm{\mu m}$), ``$32$ and $16$ pixels'' ($\approx 6.4$ and $3.2~\mathrm{\mu m}$), and ``$16$ and $8$ pixels'' ($\approx 3.2$ and $1.6~\mathrm{\mu m}$). Velocity validation with ``auto'' velocity limits is applied, while other PIVlab settings remain as default. The background flow is subtracted, which is calculated as the ensemble-average flow at the same time frame.

\section{C. Spatial Distributions of Pairwise Correlations and Relative Kinematics}
\begin{figure}[t!]
\centering
\includegraphics[width=\hsize]{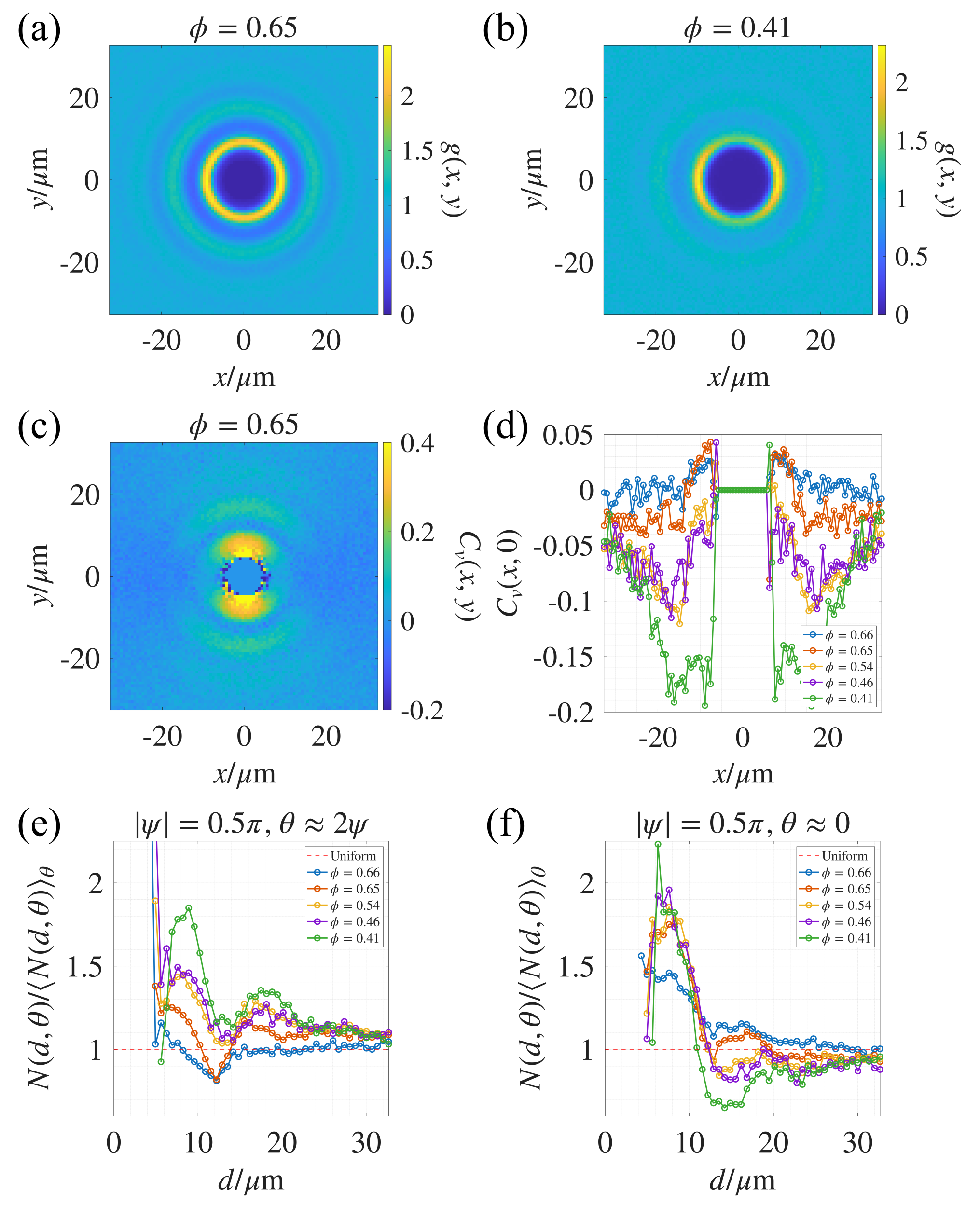}
\caption{Density- and distance-dependent competition between the two coexisting modes. (a),(b) Pair correlation $g(x,y)$ at area fraction (a) $\phi = 0.65$ and (b) $\phi = 0.41$. (c) Velocity correlation $C_v(x,y)$ at area fraction $\phi = 0.65$. To suppress small-$d$ sampling noise, the colorbar scale is adjusted to encompass $99.4\%$ of the data distribution.  (d) Lateral profiles $C_v(x,0)$, obtained by averaging $C_v(x,y)$ within $|y| < 1~\mathrm{\mu m}$. To eliminate similar sparse-sampling artifacts, noisy data at small distances with pair counts per grid element $N^{\mathrm{count}} < 50$ are truncated to zero. The mean pair counts per grid cell $\bar{N}^{\mathrm{count}}$ at decreasing $\phi$ are $2528$, $2470$, $1721$, $598$, and $964$, respectively. (e),(f) Normalized lateral pair distribution $N(d,\theta)/\left\langle N(d,\theta)\right\rangle_{\theta}$ for (e) $\theta = 2\psi$ mode ($\left\vert\psi\right\vert\in [0.45\pi, 0.55\pi]$, $\theta - 2\psi\in [-0.05\pi, 0.05\pi] \pm 2n\pi, \ n=0,\pm 1$) and (f) $\theta = 0$ mode ($\left\vert\psi\right\vert\in [0.45\pi, 0.55\pi]$, $\theta \in [-0.05\pi, 0.05\pi]$).}
\label{figS1}
\end{figure}

We first evaluate the pairwise correlation $g(x,y)$ to quantify the spatial organization of the suspension [Fig.~\ref{figS1}(a)~and~\ref{figS1}(b)]. We define $\vec{\xi}_{ij}(x,y,t)$ and pair correlation $g(x,y)$ as:
\begin{flalign}
\vec{\xi}_{ij}(x,y,t) &= x\hat{x}_i(t)+y\hat{y}_i(t)-\vec{r}_{ji}(t),\\
g(x,y) &= \frac{1}{\rho}\left\langle \sum_{j\neq i} \delta\left[ \vec{\xi}_{ij}(x,y,t)\right]\right\rangle_{i,t},
\end{flalign}
where $\delta[\vec{\xi}_{ij}(x,y,t)]$ is the Dirac delta function, $x, y$ are defined in the local body coordinate system of particle $i$ [see Fig.~\ref{M-fig2}(a)~and~\ref{M-fig2}(c) for the definition of the local body coordinate system] with the unit coordinate vectors $\hat{x}_i(t)$ and $\hat{y}_i(t)$ at time $t$, the vector $\vec{r}_{ji}(t) = \vec{r}_j(t)-\vec{r}_i(t)$ denotes the position of particle $j$ relative to particle $i$ at time $t$, $\left\langle\cdots\right\rangle_{i,t}$ represents the average over all particles $i$ and all time frames $t$, and $\rho$ is the number density. In the dense suspension ($\phi = 0.65$), the pair correlation $g(x,y)$ exhibits an isotropic structure characterized by concentric circular rings [Fig.~\ref{figS1}(a)]. By contrast, at the lower area fraction ($\phi = 0.41$), the pair correlation $g(x,y)$ becomes anisotropic, featuring an enhanced probability along the lateral direction [Fig.~\ref{figS1}(b)].

To characterize the kinetic correlation of cell pairs, we analyze the velocity correlation $C_v(x,y)$ [Fig.~\ref{figS1}(c)] and quantify the competitions of two modes by comparing the lateral profiles of velocity correlation $C_v(x,0)$ at various number densities [Fig.~\ref{figS1}(d)]. We define the velocity correlation $C_v(x,y)$ as:
\begin{flalign}
    C_v(x,y) = 
    \frac{\left\langle\sum_{j\neq i}[\vec{v}_i(t)\cdot\vec{v}_j(t)]\delta\left[ \vec{\xi}_{ij}(x,y,t)\right]\right\rangle_{i,t}}{\left\langle \vert\vec{v}_i(t)\vert^2\right\rangle_{i,t}\cdot\left\langle \sum_{j\neq i} \delta\left[\vec{\xi}_{ij}(x,y,t)\right]\right\rangle_{i,t}},
    \label{eqS2}
\end{flalign}
where $\vec{v}_i(t), \vec{v}_j(t)$ are the velocities of particle $i$ and $j$ at time $t$. In the dense suspension ($\phi=0.65$), the velocity correlation $C_v(x,y)$ exhibits strong longitudinal and weak lateral positive correlations [Fig.~\ref{figS1}(c)]. Along the lateral direction of the cell ($y=0$), the nearest-neighbor extremum of $C_v$ flips from negative to positive as $\phi$ increases from $0.41$ to $0.65$, marking a direct transition from ($\theta=2\psi$)-dominated to ($\theta=0$)-dominated lateral velocity correlations [Fig.~\ref{figS1}(d)].

To unravel the specific mode contributions underlying this lateral correlation flip [Fig.~\ref{figS1}(d)], a local population analysis of the normalized lateral pair distribution $N(d,\theta)/\langle N(d,\theta)\rangle$ is performed around $\psi = \pm \pi/2$ regions [Figs.~\ref{figS1}(e)~and~\ref{figS1}(f)]. It confirms that the $\theta=0$ contribution is strictly localized in the near field, peaking at $d \approx 8~ \mathrm{\mu m}$ before decaying to a uniform state at $d \approx 12~\mathrm{\mu m}$, with its relative proportion remaining insensitive to $\phi$ [Fig.~\ref{figS1}(e)]. In contrast, while the $\theta=2\psi$ mode exhibits a much longer spatial decay length, its overall amplitude is severely suppressed as the number density increases [Fig.~\ref{figS1}(f)].

\begin{figure}[t!]
\centering
\includegraphics[width=\hsize]{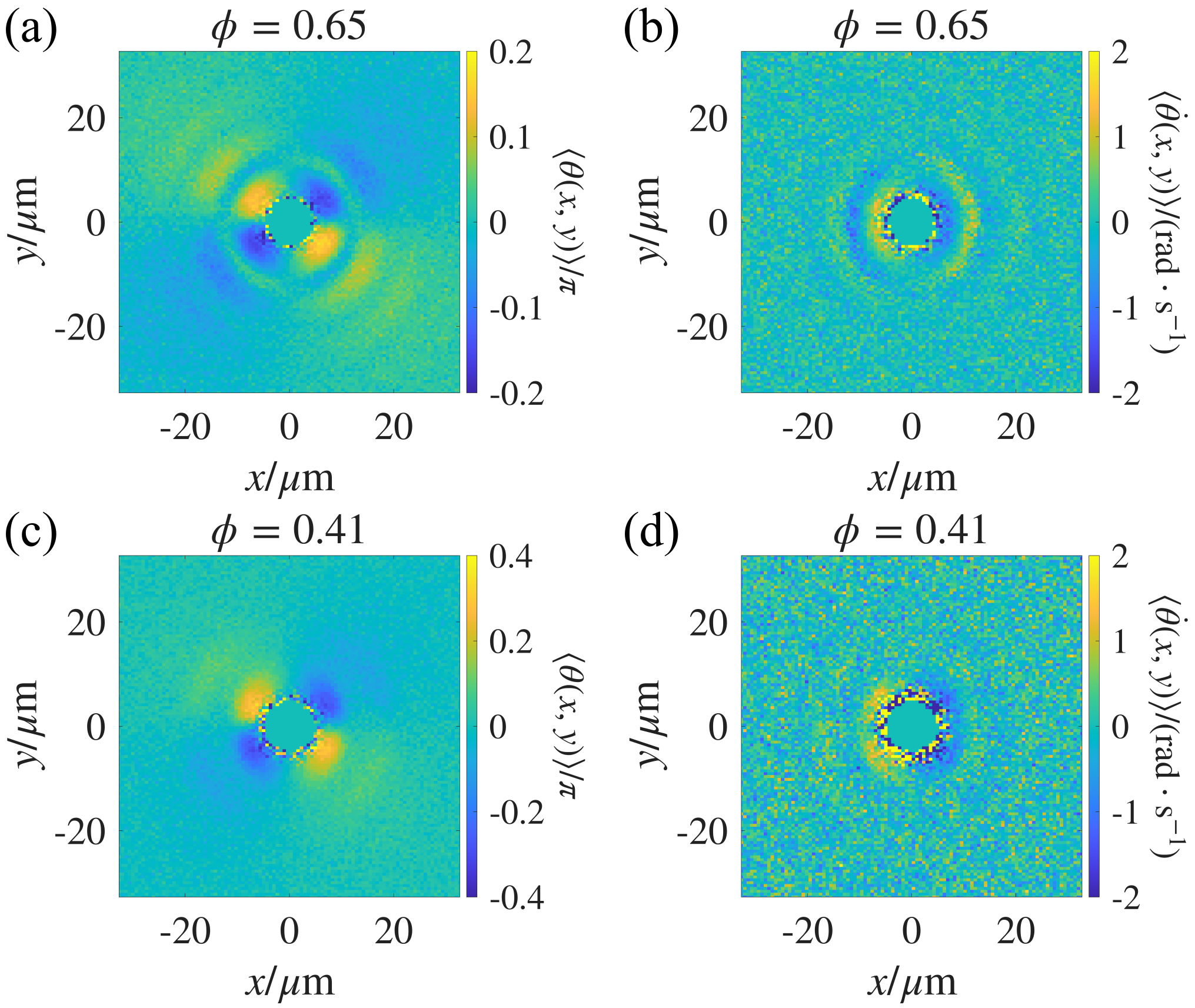}
\caption{Ensemble-averaged spatial distributions of the relative velocity angle $\langle\theta(x,y)\rangle$ and its angular velocity $\langle\dot{\theta}(x,y)\rangle$ for cell pairs. The quantities are mapped in the body-fixed frame of a reference cell. (a),(b) Results for area fraction $\phi = 0.65$. (c),(d) Results for $\phi = 0.41$. To mitigate statistical artifacts arising from sparse sampling of small-sized particle pairs with small distance $d=\sqrt{x^2+y^2}$, colorbar scales are set to encompass over $98.5\%$ of the data distribution.}
\label{figS2_oldS4}
\end{figure}

In addition to characterizing the translational correlations, we also evaluate the ensemble-averaged spatial distribution of the relative velocity angle $\langle\theta(x,y)\rangle$ [Figs.~\ref{figS2_oldS4}(a)~and~\ref{figS2_oldS4}(c)] and its angular velocity $\langle\dot{\theta}(x,y)\rangle$ [Figs.~\ref{figS2_oldS4}(b)~and~\ref{figS2_oldS4}(d)], which provide insights into the angular kinematic correlations within the suspension. The $\langle\theta(x,y)\rangle$ and $\langle\dot{\theta}(x,y)\rangle$ are defined as:
\begin{flalign}
\langle\theta(x,y)\rangle = \frac{\left\langle\sum_{j\neq i}[\theta_j(t)-\theta_i(t)]\delta\left[ \vec{\xi}_{ij}(x,y,t)\right]\right\rangle_{i,t}}{\left\langle \sum_{j\neq i} \delta\left[ \vec{\xi}_{ij}(x,y,t)\right]\right\rangle_{i,t}},    \\
\langle\dot{\theta}(x,y)\rangle = \frac{\left\langle\sum_{j\neq i}[\dot{\theta}_j(t)-\dot{\theta}_i(t)]\delta\left[ \vec{\xi}_{ij}(x,y,t)\right]\right\rangle_{i,t}}{\left\langle \sum_{j\neq i} \delta\left[ \vec{\xi}_{ij}(x,y,t)\right]\right\rangle_{i,t}},  \label{eqS4}
\end{flalign}
where $\theta_j(t)-\theta_i(t)$ and $\dot{\theta}_j(t)-\dot{\theta}_i(t)$ are respectively the relative velocity angle and the relative angular velocity of particle $j$ with respect to particle $i$ at time $t$. Regarding radial profiles, both $\langle\theta(x,y)\rangle$ and $\langle\dot{\theta}(x,y)\rangle$ exhibit radial oscillations, characterized by periodic alternating positive and negative signs as the center-of-mass distance $d$ increases. Regarding angular distributions, taking the distance range of $d \in [4~\mu\mathrm{m}, 9~\mu\mathrm{m}]$ as an example, $\langle\theta(x,y)\rangle$ exhibits a two-fold angular symmetry [Fig.~\ref{figS2_oldS4}(a)~and~\ref{figS2_oldS4}(c)], and the relative angular velocity $\langle\dot{\theta}(x,y)\rangle$ presents a one-fold angular symmetry [Fig.~\ref{figS2_oldS4}(b)~and~\ref{figS2_oldS4}(d)].

\section{D. Model of Flow Field Fitting}

Considering a quasi-2D space confined by two no-slip boundaries, the 3D flow field $\widetilde{\boldsymbol{u}}$ satisfies the Stokes equation $\widetilde{\nabla}^2\widetilde{\boldsymbol{u}} = \widetilde{\nabla}\widetilde{p}/\eta$, the incompressibility condition $\widetilde{\nabla}\cdot\widetilde{\boldsymbol{u}} = 0$, and the no-slip boundary condition $\widetilde{\boldsymbol{u}}(x,y,z=\pm H/2)=0$, where $\widetilde{\nabla} = (\partial_x,\partial_y,\partial_z)^{\mathrm{T}}$ is the 3D Del operator, and $\eta$ and $\widetilde{p}$ are viscosity and pressure, respectively. With the thin-film approximation in the far field for the quasi-2D space, a parabolic profile is assumed to describe the flow field $\widetilde{\boldsymbol{u}} = (u_x(x,y)f(z), u_y(x,y)f(z),0)^{\mathrm{T}}$ where $f(z) = \frac{3}{2}-\frac{6z^2}{H^2}$ satisfies $f(\pm H/2) = 0$. Taking the 2D depth-average operation on the flow field, the $z$-component $\partial_z^2$ of 3D Laplacian $\widetilde{\nabla}^2$ acts as a friction term $(\frac{1}{H}\int_{-H/2}^{H/2}\partial_z^2f(z)\mathrm{d}z)\boldsymbol{u} = -\alpha^2\boldsymbol{u}$ where $\alpha = \sqrt{12}/H$. Therefore, after including the effect of two no-slip boundaries, the 2D depth-average Stokes equation takes the mathematic form of the Brinkman equation \cite{Nagel2015,Leiderman2016}:
\begin{gather}
    \nabla^2\boldsymbol{u} - \alpha^2\boldsymbol{u} = \frac{1}{\eta}\nabla p,
\end{gather}
with incompressibility condition $\nabla\cdot \boldsymbol{u} = 0$, where $\nabla = (\partial_x, \partial_y)^{\mathrm{T}}$ is the 2D Del operator, and $\boldsymbol{u} = (u_x(x,y), u_y(x,y))^{\mathrm{T}}$ is the 2D depth-averaged flow. 

To compute the 2D singular Brinkmanlet, we consider a point force $\boldsymbol{f}\delta(\boldsymbol{r})$ at $\boldsymbol{r}=0$, then the equations become $\nabla^2\boldsymbol{u}-\alpha^2\boldsymbol{u}=\nabla p/\eta - \boldsymbol{f}\delta(\boldsymbol{r})$ and $\nabla\cdot\boldsymbol{u}=0$, where $ \boldsymbol{f} = \boldsymbol{F}/(\eta H)$, and $\delta(\boldsymbol{r})$ is the Dirac delta function satisfying $\int_{-\infty}^{\infty}\int_{-\infty}^{\infty}\delta(\boldsymbol{r}) \mathrm{d}^2 \boldsymbol{r} = 1$. The solution of the 2D singular Brinkmanlet \cite{Leiderman2016} is:
\begin{flalign}
    \boldsymbol{u}(\boldsymbol{r}) &= \left[ H_1(r)\mathbbm{1}+r^2H_2(r)\hat{\boldsymbol{r}}\hat{\boldsymbol{r}}\right]\cdot \boldsymbol{f},\\
    H_1(r) &= \frac{1}{2\pi}\left[ K_0(\alpha r) + \frac{K_1(\alpha r)}{\alpha r} - \frac{1}{\alpha^2 r^2}\right], \\
    H_2(r) &= \frac{1}{2\pi r^2}\left[ -K_0(\alpha r) - \frac{2K_1(\alpha r)}{\alpha r} + \frac{2}{\alpha^2 r^2}\right],
\end{flalign}
where $\mathbbm{1}$ is the identity matrix, and $K_0, K_1$ are the modified Bessel functions of the second kind. The asymptotic properties of the 2D singular Brinkmanlet are:
\begin{gather}
   \alpha r\rightarrow0: \ \boldsymbol{u}(\boldsymbol{r})\sim-\frac{1}{2\pi\eta H}\left\{ \left[\frac{1}{2}\mathrm{ln}\left(\frac{\alpha r}{2}\right) + \gamma\right] \mathbbm{1} - \frac{1}{2}\hat{\boldsymbol{r}}\hat{\boldsymbol{r}}\right\} \cdot \boldsymbol{F},\\
    \alpha r\rightarrow \infty: \ \boldsymbol{u}(\boldsymbol{r})\sim -\frac{1}{2\pi\eta H}\frac{1}{\alpha^2r^2}\left( \mathbbm{1} - 2\hat{\boldsymbol{r}}\hat{\boldsymbol{r}}\right) \cdot \boldsymbol{F},  
\end{gather}
where $\gamma \approx 0.577$ is the Euler constant. In the far field, the 2D singular Brinkmanlet has the same Oseen tensor $(\mathbbm{1}-2\hat{\boldsymbol{r}}\hat{\boldsymbol{r}})/r^2$ as the 2D source dipole $\boldsymbol{u}_{\mathrm{sd}}$ of strength $\boldsymbol{I}_{\mathrm{sd}}$ \cite{Liron1976}:
\begin{flalign}
    \boldsymbol{u}_{\mathrm{sd}}(\boldsymbol{r}) = -\frac{1}{2\pi r^2}(\mathbbm{1}-2\hat{\boldsymbol{r}}\hat{\boldsymbol{r}})\cdot\boldsymbol{I}_{\mathrm{sd}},
\end{flalign}
which is an irrotational potential flow: $\nabla\times\boldsymbol{u}_{\mathrm{sd}} = 0$. This is because $\vert\nabla^2\boldsymbol{u}\vert \ll \vert \alpha^2\boldsymbol{u}\vert$ for $\alpha r\gg 1$ in the far field, which makes the Brinkman flow become a potential flow following Darcy's law \cite{Batchelor1967}: $\boldsymbol{u}=-\nabla p/(\eta \alpha^2)$ for $\alpha r\gg1$. Importantly, the Oseen tensor $(\mathbbm{1}-2\hat{\boldsymbol{r}}\hat{\boldsymbol{r}})/r^2$ strictly satisfies $\theta = 2\psi$ under our angular definitions.

To make the fitting smoother, we reduce the effect of singularities by using 2D regularized Brinkmanlet. We regard the spatial distribution of force $\boldsymbol{f}\phi_{\delta}(\boldsymbol{r})$ as a blob rather than a point: $\phi_{\delta}(\boldsymbol{r}) = 3\delta^3/\left[ 2\pi(r^2+\delta^2)^{5/2}\right]$ satisfying $2\pi\int_{0}^{\infty}r\phi_{\delta}\mathrm{d}r=1$. Then the 2D regularized Brinkmanlet \cite{Leiderman2016} is:
\begin{flalign}
    \boldsymbol{u}(\boldsymbol{r}) &= \left[ H_1^{\delta}(R, \alpha)\mathbbm{1}+r^2H_2^{\delta}(R,\alpha)\hat{\boldsymbol{r}}\hat{\boldsymbol{r}}\right]\cdot \boldsymbol{f}, \\
    H_1^{\delta}(R, \alpha) &= \frac{1}{2\pi}\left[\frac{r^2}{R^2}K_0(\alpha R) + \frac{r^2-\delta^2}{\alpha R^3}K_1(\alpha R) - \frac{r^2-\delta^2}{\alpha^2R^4}\right], \\
    H_2^{\delta}(R,\alpha) &= \frac{1}{2\pi R^2}\left[ -K_0(\alpha R) - \frac{2K_1(\alpha R)}{\alpha R} + \frac{2}{\alpha^2R^2}\right],
\end{flalign}
where $R = \sqrt{r^2 + \delta^2}$.

\section{E. Procedure and Results of Flow Field Fitting}

\begin{table*}[!htpb]
\caption{\label{tab:fit_results} Summary of the fitting models, parameter range, and results.}

\begin{ruledtabular} 
\begin{tabular*}{\textwidth}{@{\extracolsep{\fill}} l c c c c c}
Model & Fitting Range & $H\in[10~\mathrm{\mu m}, 50~\mathrm{\mu m}]$ & $F\in[0~\mathrm{pN}, 20~\mathrm{pN}]$ & $\delta\in[0~\mathrm{\mu m}, 8~\mathrm{\mu m}]$ & $F^{\prime}\in[0~\mathrm{pN}, 10~\mathrm{pN}]$\\
\colrule 
Unbalanced ($F > 2F^{\prime}$) & $r\in[5~\mathrm{\mu m}, 40~\mathrm{\mu m}]$ & $14.39~\mathrm{\mu m}$ & $11.12~\mathrm{pN}$ & $7.26~\mathrm{\mu m}$ & $3.36~\mathrm{pN}$  \\
Balanced ($F = 2F^{\prime}$) & $r\in[20~\mathrm{\mu m}, 40~\mathrm{\mu m}]$ & $22.15~\mathrm{\mu m}$ & $7.44~\mathrm{pN}$ & $7.99~\mathrm{\mu m}$ & $F^{\prime} = F/2=3.72~\mathrm{pN}$ \\
Balanced ($F = 2F^{\prime}$) & $r\in[5~\mathrm{\mu m}, 40~\mathrm{\mu m}]$ & $40.01\mathrm{\mu m}$ & $15.88~\mathrm{pN}$ & $5.51~\mathrm{\mu m}$ & $F^{\prime} = F/2 = 7.94~\mathrm{pN}$ 
\end{tabular*}
\end{ruledtabular}
\vspace{0.1cm}
\begin{ruledtabular}
\begin{tabular*}{\textwidth}{@{\extracolsep{\fill}} l c c c c c }
Model & Fitting Range & $\delta^{\prime}\in[0~\mathrm{\mu m}, 12~\mathrm{\mu m}]$ & $r_f\in[5~\mathrm{\mu m}, 20~\mathrm{\mu m}]$ & $\beta\in[0,\pi/2]$ & $I_{\mathrm{sd}}\in[0, 20000~\mathrm{\mu m^3/s}]$ \\
\colrule
Unbalanced ($F > 2F^{\prime}$) & $r\in[5~\mathrm{\mu m}, 40~\mathrm{\mu m}]$ & $10.32~\mathrm{\mu m}$ & $11.24~\mathrm{\mu m}$ & $0.26\pi$ & $7.31~\mathrm{\mu m^3/s}$ \\
Balanced ($F = 2F^{\prime}$) & $r\in[20~\mathrm{\mu m}, 40~\mathrm{\mu m}]$ & $10.19~\mathrm{\mu m}$ & $7.19~\mathrm{\mu m}$ & $0.30\pi$ & $4543.55~\mathrm{\mu m^3/s}$ \\
Balanced ($F = 2F^{\prime}$) & $r\in[5~\mathrm{\mu m}, 40~\mathrm{\mu m}]$ & $12.00~\mathrm{\mu m}$ & $5.00~\mathrm{\mu m}$ & $0.14\pi$ & $0.09\mathrm{\mu m^3/s}$
\end{tabular*}
\end{ruledtabular}

\end{table*}

\begin{figure}[t!]
\centering
\includegraphics[width=\hsize]{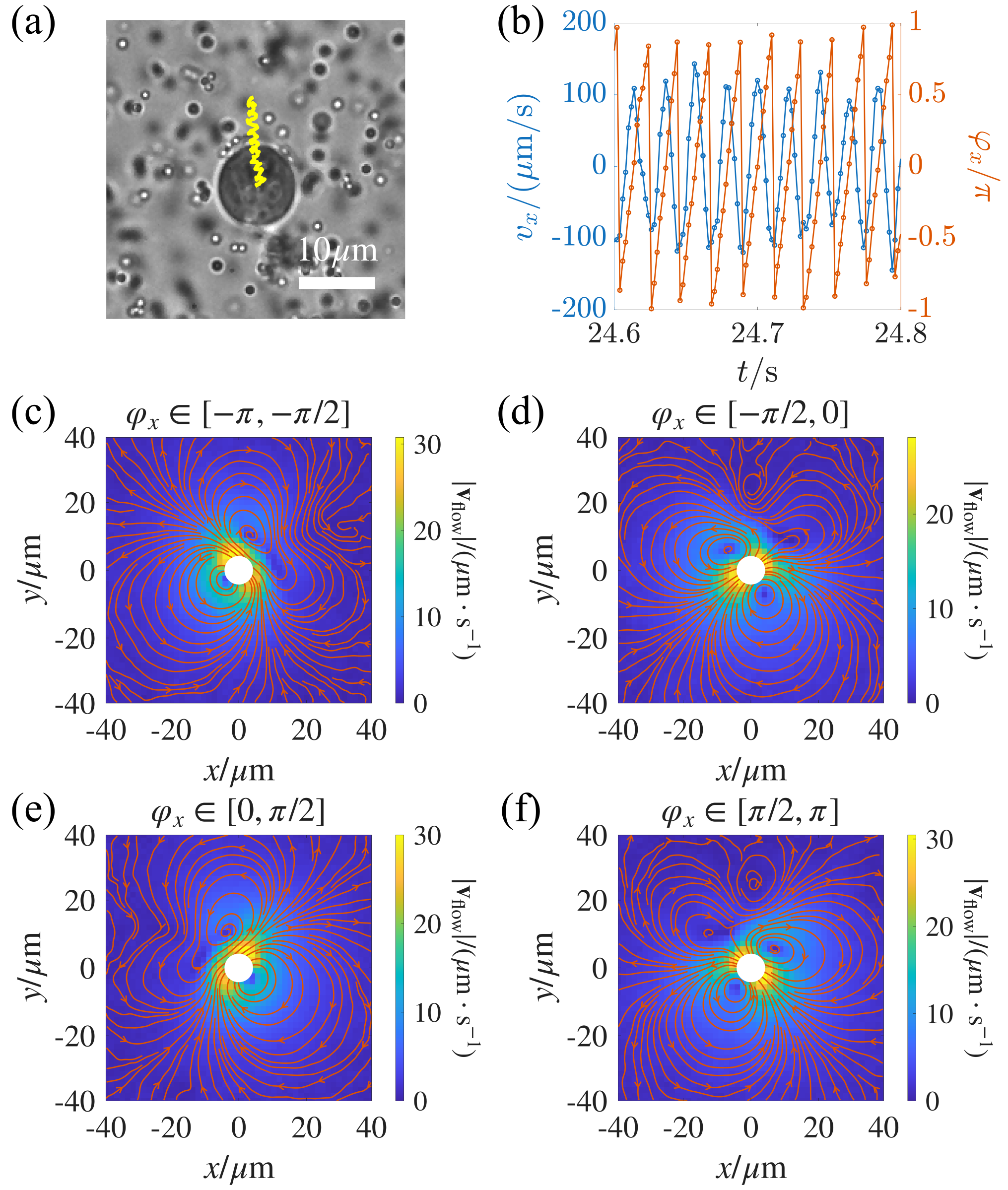}
\caption{Oscillatory flow fields induced by a single \textit{C. reinhardtii} cell (wobbler) during one self-propulsion period. (a) Experimental snapshot and oscillatory trajectory (yellow curve) (in the part $0.2~\mathrm{s}$) of a single cell in the suspension mixed with $1$-$\mathrm{\mu m}$-diameter tracer beads. (b) Time evolution of the lateral smoothed velocity component $v_x\propto \mathrm{cos}(2\pi ft + \varphi_x)$ (mean frequency $\bar{f} \approx 50~\mathrm{Hz}$, smoothing window: $0.01~\mathrm{s}$) and its corresponding phase $\varphi_x$ extracted via Hilbert transform. Coordinates follow the local body frame in \figpref{M-fig2}{a}. (c)--(f) Phase-resolved mean flow fields during specific stroke intervals: (c) $\varphi_x\in [-\pi, -\pi/2]$; (d) $\varphi_x\in [-\pi/2, 0]$; (e) $\varphi_x\in [0, \pi/2]$; (f) $\varphi_x\in [\pi/2, \pi]$. Diagram conventions in (c)--(f) follow \figpref{M-fig2}{a}.}
\label{figS3_oldS2}
\end{figure}

\begin{figure}[t!]
\centering
\includegraphics[width=\hsize]{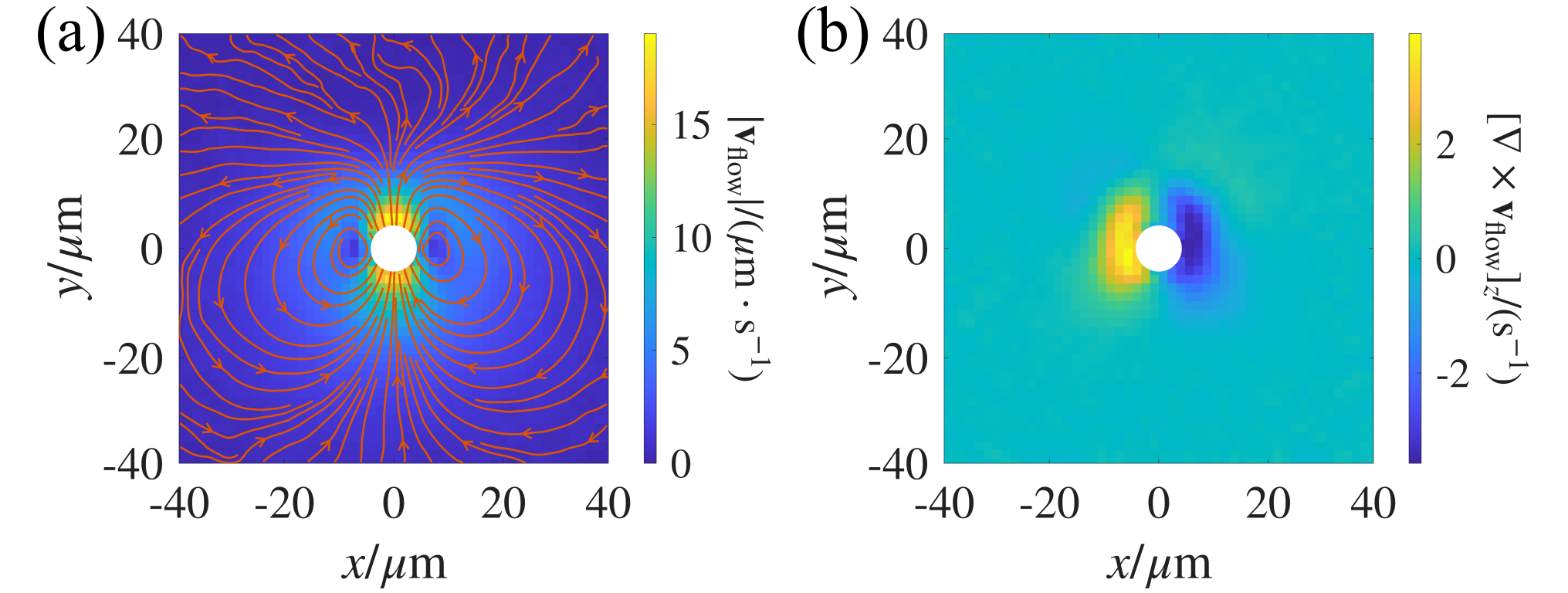}
\caption{Mean flow field around a single \textit{C. reinhardtii} cell. (a) Ensemble-averaged flow field. (b) Spatial distribution of the mean flow's out-of-plane vorticity, $\left[ \nabla\times \boldsymbol{v}_{\mathrm{flow}}\right]_z$, highlighting strong primary lateral vortices adjacent to the cell body and weaker secondary vortices flanking the two front flagella.}
\label{figS4_oldS3}
\end{figure}

To fit the experimental flow field (Fig.~\ref{figS4_oldS3}; see Fig.~\ref{figS3_oldS2} for phase-resolved flows), we utilize a 2D source dipole at the body center $(0,0)$, and three regularized Brinkmanlets ($F$ at body center $(0,0)$, two $F^{\prime}$ at two flagella $(\pm x_f, y_f)$ where $x_f = r_f\cos\beta, \ y_f = r_f\sin\beta$). For the unbalanced fitting ($F>2F^{\prime}$), we have $8$ free-fitting parameters ($H$, $F$, $\delta$, $F^{\prime}$, $\delta^{\prime}$, $r_f$, $\beta$, $I_{sd}$), to fit the flow at both near field and far field ($r\in[5~\mathrm{\mu m}, 40~\mathrm{\mu m}]$). For the balanced fitting, we set the restriction $F = 2F^{\prime}$ so that we have $7$ free-fitting parameters: ($H$, $F$, $\delta$, $\delta^{\prime}$, $r_f$, $\beta$, $I_{sd}$), to fit the flow only at far field ($r\in[20~\mathrm{\mu m}, 40~\mathrm{\mu m}]$). We use MATLAB function \textit{fmincon} to find minimum of constrained nonlinear multivariable function, and minimize the relative error defined as $\varepsilon(\left\{ x_i,y_j\right\}) = \sum_{i,j}\left( \boldsymbol{u}_{\mathrm{exp}}(x_i,y_j)-\boldsymbol{u}_{\mathrm{fit}}(x_i,y_j)\right)^2/\sum_{i,j}\boldsymbol{u}_{\mathrm{exp}}^2(x_i,y_j)$. \cite{Jeanneret2019} The summary of the fitting models, the fitting range of parameters, and the fitting results of parameters is shown in Table~\ref{tab:fit_results}.

\begin{figure}[t!]
\centering
\includegraphics[width=\hsize]{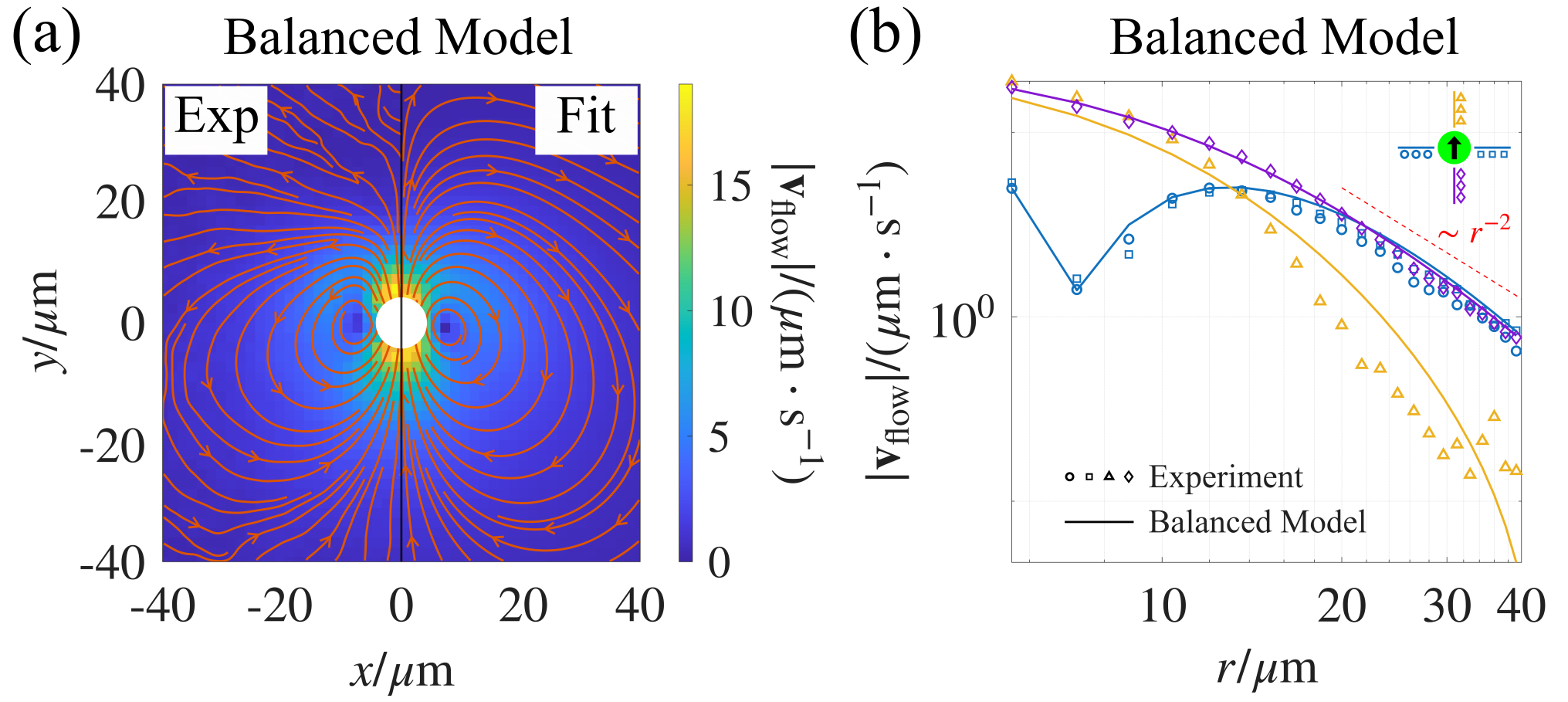}
\caption{Single-cell flow fitting using the balanced model with fitting region $[5~\mathrm{\mu m}, 40~\mathrm{\mu m}]$. (a) Left panel: the experimental flow around a single cell; Right panel: the fitting flow of the balanced model ($F=2F^{\prime}$). Notation in (a) follows that of Fig.~\ref{M-fig2}(a). (b) Experimental measurements (symbols) and balanced model fits (lines) of flow profiles along four directions extracted from (a), indicated by separate colors and symbols in the inset.}
\label{figS}
\end{figure}

Figure~\ref{M-fig2} compares the fitting results of the balanced model (fitting range: $[20~\mathrm{\mu m}, 40~\mathrm{\mu m}]$) and the unbalanced model (fitting range: $[5~\mathrm{\mu m}, 40~\mathrm{\mu m}]$). For a fair comparison, we also perform a fitting by the balanced model with fitting range $[5~\mathrm{\mu m}, 40~\mathrm{\mu m}]$ (Fig.~\ref{figS}~and~Table~\ref{tab:fit_results}). In this case, we find the fitted magnitude of the 2D source dipole nearly vanishes (Table~\ref{tab:fit_results}), causing the loss of the $\theta= 2\psi$ signature [Fig.~\ref{figS}(a)] and $\sim r^{-2}$ decay law [Fig.~\ref{figS}(b)] in the far field, farther than the range of $r$ shown in Fig.~\ref{figS}. This shows that the fitting results of the balanced model (fitting range: $[5~\mathrm{\mu m}, 40~\mathrm{\mu m}]$) is not only physically inconsistent but also fails to provide a good approximation for describing the flow field.

For fitting the simulated flow field [Fig.~\ref{M-fig3}(b) in the main text], we employ a combination of a single regularized Brinkmanlet and a source dipole, both centered at the particle center. We fix the system hight at $H=15$, leaving three free-fitting parameters, $F$, $\delta$, and $I_{\mathrm{sd}}$. The best-fit values are $F=0.179$, $\delta=8.68$, and $I_{\mathrm{sd}}=0.0325$.

\section{F. Simulation Method}
In our hydrodynamic simulations, we employ the smoothed profile method (SPM) \cite{Nakayama2005, Yamamoto2021} to implement hydrodynamic interactions among the particles. In this method, the particles are represented by a scalar field variable called a profile function $\Phi(\boldsymbol{r},t)$. In the present work, we adopt the following form
\begin{align}
    \Phi(\boldsymbol{r},t) &= \sum_{i=1}^N \Phi_i(\boldsymbol{r},t), \\
    \Phi_i(\boldsymbol{r},t) &= s(r_0 - |\boldsymbol{r}-\boldsymbol{R}_i(t)|), \\
    s(x) &= 
    \begin{cases}
        0, & x <- \dfrac{\xi}{2}, \\
        \dfrac{1}{2} \left( \sin{\dfrac{\pi x}{\xi}} + 1 \right), & |x|<\dfrac{\xi}{2}, \\
        1, & x > \dfrac{\xi}{2},
    \end{cases}
\end{align}
where $r_0$ is the radius of particles, $\boldsymbol{R}_i(t)$ is the position vector of the $i$th particle, and $\xi$ denotes the interfacial thickness. In the SPM, the total velocity field is defined as
\begin{align}
    \boldsymbol{v}(\boldsymbol{r},t) = (1-\Phi) \boldsymbol{v}_{\rm f}(\boldsymbol{r},t) + \Phi \boldsymbol{v}_{\rm p}(\boldsymbol{r},t),
\end{align}
where $\boldsymbol{v}_{\rm f}$ and $\boldsymbol{v}_{\rm p}$ are the host fluid and particle velocity fields, respectively. The velocity field of particles is defined as
\begin{align}
    \Phi \boldsymbol{v}_{\rm p}(\boldsymbol{r},t) = \sum_{i=1}^N \Phi_i(\boldsymbol{r},t) \left( \boldsymbol{V}_i(t) + \boldsymbol{\Omega}_i(t) \times [\boldsymbol{r} - \boldsymbol{R}_i(t)] \right),
\end{align}
where $\boldsymbol{V}_i(t)$ and $\boldsymbol{\Omega}_i(t)$ are the translational and angular velocities of the $i$th particle, respectively. The total velocity field $\boldsymbol{v}(\boldsymbol{r},t)$ is governed by the following Navier-Stokes equation and the incompressible condition:
\begin{align}
    \rho \dfrac{\partial \boldsymbol{v}}{\partial t} &= \boldsymbol{\nabla} \cdot \left( -\rho \boldsymbol{v} \boldsymbol{v} + \boldsymbol{\sigma} \right) + \boldsymbol{f}_{\rm p}, \\
    \boldsymbol{\sigma} &= -p \mathbbm{1} + \eta [\boldsymbol{\nabla}\boldsymbol{v} + (\boldsymbol{\nabla}\boldsymbol{v})^{\rm T}], \\
    \boldsymbol{\nabla} \cdot \boldsymbol{v} &= 0,
\end{align}
where $p$ is the pressure field and the body-force density $\boldsymbol{f}_{\rm p}$ is introduced to impose the rigidity constraint by enforcing the rigid-body velocity field in the particle domain on the total velocity field, while maintaining the incompressibility condition.

The equations of motion for the $N$-dragged particles suspended in a host fluid are
\begin{flalign}
    M_i \dfrac{d \boldsymbol{V}_i}{d t} &= \boldsymbol{F}^{\rm H}_i + \boldsymbol{F}^{\rm A}_i + \boldsymbol{F}^{\rm C}_i, \\
    \boldsymbol{I}_i \cdot \dfrac{d \boldsymbol{\Omega}_i}{d t} &= \boldsymbol{N}^{\rm H}_i + \boldsymbol{N}^{\rm R}_i,
\end{flalign}
where $M_i$, and $\boldsymbol{I}_i$ are the mass and the moment-of-inertia tensor of the $i$th particle, respectively.
$\boldsymbol{F}^{\rm H}_i$($\boldsymbol{N}^{\rm H}_i$), $\boldsymbol{F}^{\rm A}_i$, and $\boldsymbol{F}^{\rm C}_i$ are the hydrodynamic force(torque), the active force and the direct force arising from particle–particle and particle–wall interactions, respectively. The hydrodynamic force $\boldsymbol{F}^{\rm H}_i$ and torque $\boldsymbol{N}^{\rm H}_i$ are determined from the corresponding momentum exchange between the fluid and the particle, following Refs.~\cite{Nakayama2005, Yamamoto2021}. In the dragged-particle model considered here, the active force $\boldsymbol{F}^{\rm A}_i = F_{\rm A} \hat{\boldsymbol{n}}_i$ is applied directly to each particle, with no separate compensating force imposed on the fluid.

The direct force is defined as follows:
\begin{flalign}
    \boldsymbol{F}^{\rm C}_i = &-\sum_{j \neq i}^N \dfrac{\partial}{\partial \boldsymbol{R}_i} U_{\rm sc}(|\boldsymbol{R}_i - \boldsymbol{R}_j|)\nonumber \\
    &-\dfrac{\partial}{\partial \boldsymbol{R}_i}U_{\rm sc} \left(\dfrac{H}{2}-R_{i,z}\right) - \dfrac{\partial}{\partial \boldsymbol{R}_i} U_{\rm sc}\left(R_{i,z}+\dfrac{H}{2}\right),
\end{flalign}
where $U_{\rm sc}(r)$ is the soft-core potential:
\begin{flalign}
    U_{\rm sc}(r) = \epsilon_0 \left( \dfrac{\sigma}{r} \right)^{12}.
\end{flalign}
The first, second, and third terms describe the particle--particle interaction, the particle--top-wall interaction, and the particle--bottom-wall interaction, respectively. For the particle--wall interactions, we set $\sigma=r_0 + d_0$, where $d_0$ is a buffer length introduced to maintain a finite gap between the particle surfaces and the walls. For the particle--particle interactions, we set $\sigma = 2 r_0$, as described in the main text.

The orientation vector, $\hat{\boldsymbol{n}}_i=(\hat{n}_{i,x},\hat{n}_{i,y},0)$, is constrained to lie in the $XY$ plane. If the orientation vector is allowed to rotate freely in three dimensions, it tends to align perpendicular to the wall, causing the particle to remain immobilized at the wall. To avoid this unphysical behavior, we restrict its rotational motion to the $XY$ plane. The orientation vector is subjected to rotational diffusion driven by a random torque, $\boldsymbol{N}^{\rm R}_i=(0, 0, N^{\rm R}_{i,z})$ defined by
\begin{flalign}
\left\langle N^{\rm R}_{i,z}(t_0) \right\rangle_{t_0} = 0, \
\left\langle N^{\rm R}_{i,z}(t_0+t)N^{\rm R}_{j,z}(t_0) \right\rangle_{t_0} = \alpha^\Omega \delta_{ij}\delta(t).
\end{flalign}
In the absence of the random torque, the system develops a persistent polar order over sufficiently long times. The random torque is therefore introduced to suppress this spontaneous polarization. Here, $\alpha^\Omega$ denotes the noise strength, and $\langle\cdots\rangle_{t_0}$ represents an average over the reference time $t_0$.

In the present simulations, all quantities are expressed in terms of three fundamental units: the mesh size $\Delta$, the solvent viscosity $\eta$, and the solvent density $\rho$. The corresponding units of time and energy are $\rho \Delta^2/\eta$ and $\Delta \eta^2/\rho$, respectively. We set $\Delta=\eta=\rho=1$ and use the following parameter values: $r_0=4$, $\xi=1$, $F_{\mathrm A}=1$, $\epsilon_0 = 2$,  $\alpha^{\Omega}=2.0\times10^3$, $d_0=2$, $L=128$, and $H=15$.
 
\section{G. Lubrication Effects}

\begin{figure}[t!]
    \centering
    \includegraphics[width=\hsize]{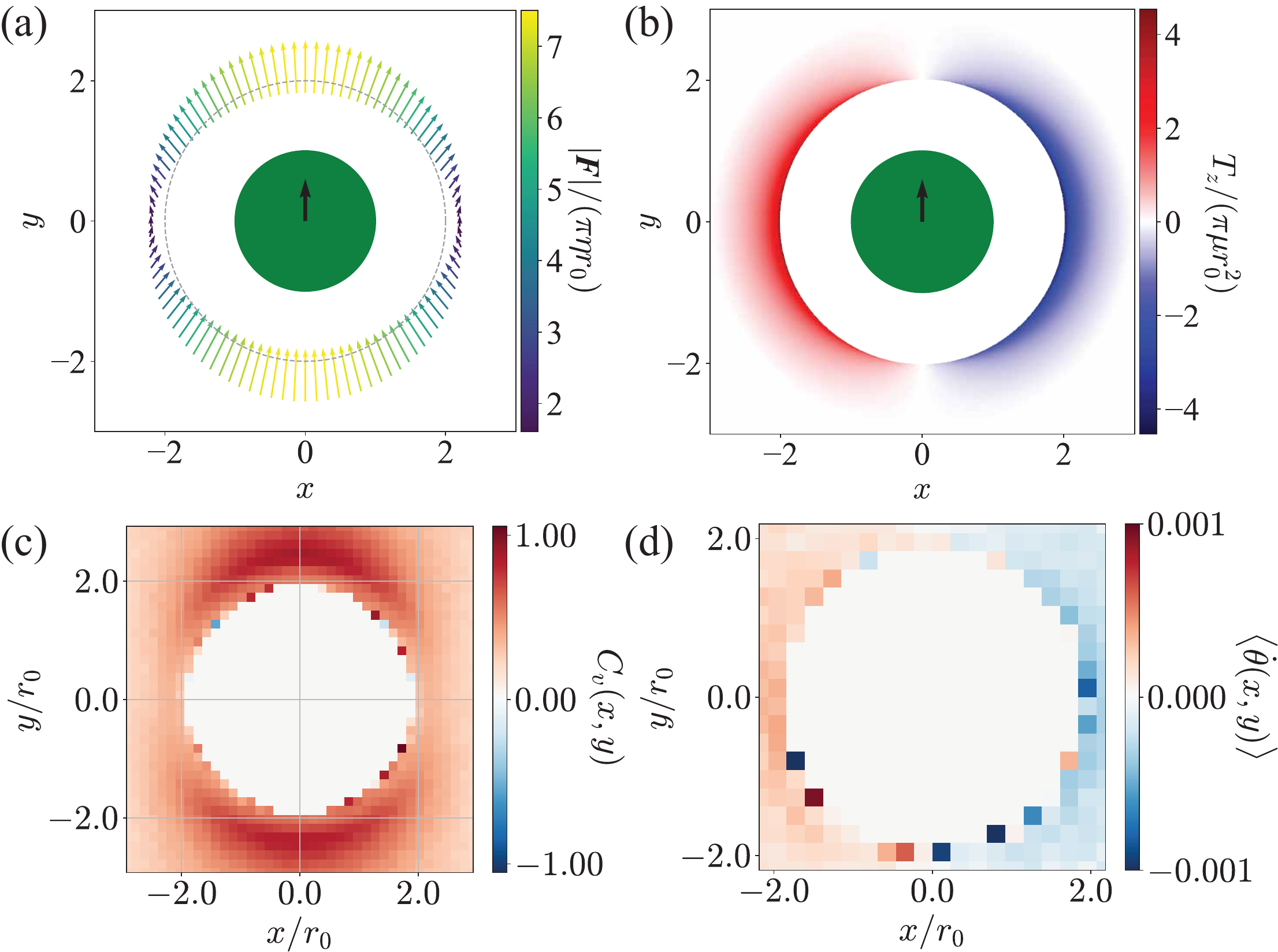}
    \caption{
    (a),(b) Lubrication-induced force and torque exerted on a stationary particle by a translating particle.
    The green disk and black arrow indicate the translating particle and its direction of motion, respectively.
    (a) Lubrication force $\boldsymbol{F}$ for a fixed gap size $r_0\epsilon$ with $\epsilon=0.2$.
    (b) Lubrication torque $T_z$ for gap sizes in the range $0.01 \leq \epsilon \leq 0.2$.
    (c) Velocity correlation $C_v(x,y)$.
    (d) Relative angular velocity $\langle \dot{\theta}(x,y) \rangle$.
    }
    \label{figS5}
\end{figure}

When two spherical particles come into close proximity, the hydrodynamic interactions are dominated by the flow in the narrow gap between them, known as lubrication \cite{kim&karrila}.
Here, we illustrate the lubrication-induced force and torque exerted on a stationary particle by a neighboring particle translating with velocity $U\hat{\boldsymbol{n}}$.
Figure~\ref{figS5}(a) shows the lubrication force for a fixed gap size $r_0\epsilon$ with $\epsilon=0.2$, whereas Fig.~\ref{figS5}(b) shows the $z$ component of the torque over the range $0.01 \leq \epsilon \leq 0.2$.

For two identical spheres of radius $r_0$, with the gap size denoted by $r_0\epsilon$, the leading-order contributions are given by
\begin{flalign}
    \boldsymbol{F} &= \pi \eta r_0 U \left[ 
    \frac{3}{2\epsilon} 
    (\hat{\boldsymbol{n}} \cdot \hat{\boldsymbol{r}}_{ij}) 
    \hat{\boldsymbol{r}}_{ij} 
    + \ln(\epsilon^{-1}) 
    \left\{
    \hat{\boldsymbol{n}}
    -(\hat{\boldsymbol{n}} \cdot \hat{\boldsymbol{r}}_{ij})
    \hat{\boldsymbol{r}}_{ij} 
    \right\} 
    \right],
    \\
    T_{z} &= \pi \eta r_0^2 \ln(\epsilon^{-1}) 
    \left( \hat{\boldsymbol{n}} \times \hat{\boldsymbol{r}}_{ij} \right)_z ,
\end{flalign}
where $\eta$, $\hat{\boldsymbol{n}}$, and $\hat{\boldsymbol{r}}_{ij}$ denote the viscosity of the suspending fluid, the unit vector along the velocity of the moving particle, and the unit vector pointing from the moving particle to the stationary particle, respectively.

We also computed the velocity correlation $C_v(x,y)$, defined in \eqref{eqS2}, and the mean relative angular velocity $\langle\dot{\theta}(x,y)\rangle$ defined in \eqref{eqS4}.

Figure~\ref{figS5}(c) shows $C_v(x,y)$ obtained from the dragged-particle simulation with $N=128$. The correlation is pronounced in front of and behind a particle, whereas it remains positive but is relatively weak at side-by-side positions. This behavior is consistent with the lubrication forces illustrated in Fig.~\ref{figS5}(a). In front of and behind a particle, the squeezing force, which diverges as $\epsilon^{-1}$, is enhanced relative to the shearing force, which diverges only logarithmically as $\log \epsilon^{-1}$.

Figure~\ref{figS5}(d) shows $\langle\dot{\theta}(x,y)\rangle$ under the same simulation conditions as those used for $C_v(x,y)$. The $j$th particle rotates counterclockwise on the left side of the $i$th particle and clockwise on its right side, consistent with the lubrication torque induced by the shearing motion between the two particles, as illustrated in Fig.~\ref{figS5}(b).

\section{H. Neutral Squirmer Model}
\begin{figure}[t!]
    \centering
    \includegraphics[width=\hsize]{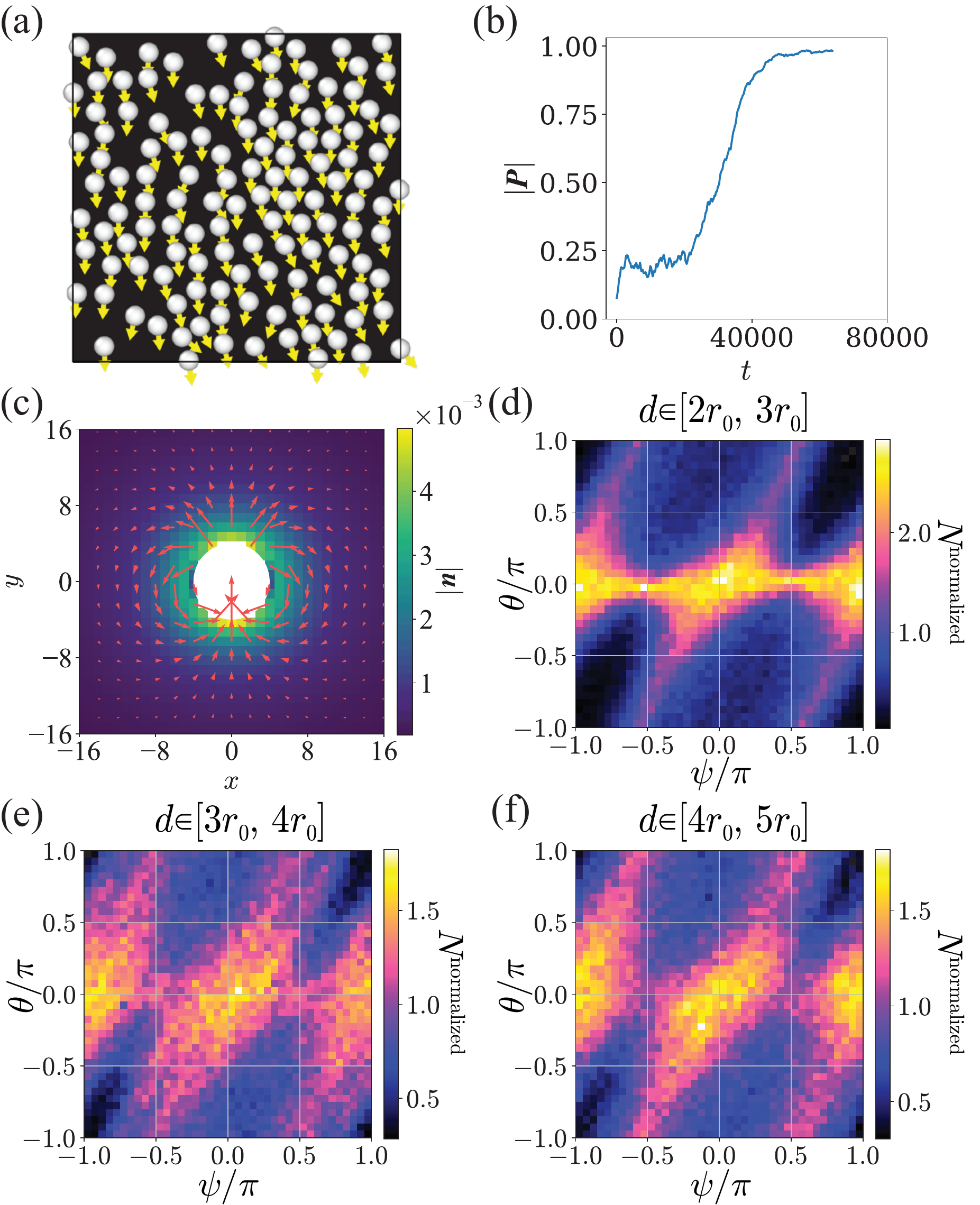}
    \caption{
    (a) Snapshot of the squirmer system, exhibiting polar order.
    (b) Time evolution of the magnitude of the polar order vector.
    (c) Depth-averaged flow field around an isolated neutral squirmer for $H=15$.
    (d)--(f) Angular pair distributions in a suspension of $N=128$ neutral squirmers for 
    (d) $d \in [2r_0, 3r_0]$, 
    (e) $d \in [3r_0, 4r_0]$, and 
    (f) $d \in [4r_0, 5r_0]$.
    }
    \label{figS6}
\end{figure}

We simulate $N$ neutral squirmers in the same simulation box as that used for the dragged-particle model described in the main text, using the smoothed profile method \cite{Molina2013}. The particle orientations are constrained to the $xy$ plane, and rotational diffusion is included to suppress polar ordering at long times.

We first examined polar ordering in the neutral-squirmer system. As shown in Fig.~\ref{figS6}(a), the system exhibits strong polar ordering, consistent with previous work \cite{Delfau2016}. Figure~\ref{figS6}(b) shows the time evolution of the magnitude of the polar order vector, $|\boldsymbol{P}(t)|$. We find that $|\boldsymbol{P}(t)|$ increases sharply after $t \sim 20000$. To eliminate the effect of polar ordering, we calculate the angular pair distributions using only data for $t \leq 24000$.

Figure~\ref{figS6}(c) shows that the flow field exhibits two vortices adjacent to the particle, in contrast to the dragged-particle case. This difference may explain why the $\theta=0$ mode at $\psi=\pm \pi/2$ at short distances is relatively weaker than that for dragged particles [Fig.~\ref{figS6}(d)]. At long distances, the $\theta=0$ mode disappears and the $\theta=2\psi$ mode emerges, as in the dragged-particle system [Figs.~\ref{figS6}(e) and \ref{figS6}(f)].

\section{I. Video Captions}

\begin{description}
\item[Video S1] Dynamics of a quasi-2D \textit{C. reinhardtii} suspension at an area fraction of $\phi = 0.65$. Scale bar: $100~\mathrm{\mu m}$. Recording frame rate: $10~\mathrm{fps}$; playback speed: $1\times$. Red vectors visualize instantaneous cell velocities, where the vector length corresponds to the distance traveled in a duration of $\tau = 1~\mathrm{s}$ at that speed. For visual clarity, velocities exceeding a threshold $v^* = 10~\mathrm{\mu m/s}$ are capped at $v^*$ for vector plotting ($\sim 4.7\%$ of total velocities). The angular pair distributions shown in Figs.~\ref{M-fig1}(e)--\ref{M-fig1}(g) were computed from this recording.

\item[Video S2] Dynamics of a quasi-2D \textit{C. reinhardtii} suspension at an area fraction of $\phi = 0.41$. Scale bar: $100~\mathrm{\mu m}$. Recording frame rate: $10~\mathrm{fps}$; playback speed: $1\times$. Red vectors visualize instantaneous cell velocities, where the vector length corresponds to the distance traveled in a duration of $\tau = 0.5~\mathrm{s}$ at that speed. For visual clarity, velocities exceeding a threshold $v^* = 30~\mathrm{\mu m/s}$ are capped at $v^*$ for vector plotting ($\sim 4.5\%$ of total velocities). The angular pair distributions shown in Figs.~\ref{M-fig1}(h)--\ref{M-fig1}(j) were computed from this recording.

\item[Video S3] Swimming of confined wobbler \textit{C. reinhardtii} cells in a suspension mixed with $1$-$\mathrm{\mu m}$-diameter tracer beads. Scale bar: $20~\mathrm{\mu m}$. Recording frame rate: $500~\mathrm{fps}$; playback speed: $0.1\times$. Blue and red vectors visualize instantaneous and smoothed velocities, respectively. Smoothed velocities were calculated using the MATLAB function \textit{movingslope} ($\text{support length} = 50\text{ frames}$, corresponding to $0.1~\mathrm{s}$). For visual clarity, vector spatial lengths are set equal to the respective cell radius to indicate motion direction, while the numbers displayed on particles denote smoothed speed magnitudes (in $\mathrm{\mu m/s}$). Yellow curves visualize trajectories in the past $0.3~\mathrm{s}$.

\item[Video S4] Dynamics of the mixtures of active \textit{C. reinhardtii} cells and passive $10$-$\mathrm{\mu m}$-diameter beads. Scale bar: $50~\mathrm{\mu m}$. Recording frame rate: $50~\mathrm{fps}$; playback speed: $0.1\times$. Red, blue, and green vectors represent active cells, passive beads, and unrecognized particles (excluded from analysis), respectively. All vectors visualize smoothed velocities (calculated via MATLAB function \textit{movingslope} with a $50$-frame window corresponding to $0.1~\mathrm{s}$) after subtracting background flow. For visual clarity, vector spatial lengths are set equal to the respective particle radius to indicate motion direction, while the numbers displayed on particles denote smoothed speed magnitudes after background flow subtraction (in $\mathrm{\mu m/s}$). The background flow with $X$-component $U^{\mathrm{bg}}$ and $Y$-component $V^{\mathrm{bg}}$ is defined as the mean smoothed velocity of passive beads, where $X, Y$ axes follow standard image pixel coordinate system. The angular pair distributions shown in Figs.~\ref{M-fig4}(b) and \ref{M-fig4}(c) were computed from this recording.

\item[Video S5]
Dragged-particle suspension simulation with $N = 224$.
The yellow arrows represent the unit orientation vectors of the particles.
The particle positions shown in the movies were recorded every 1000 time steps.
The angular pair distributions shown in Figs.~\ref{M-fig3}(c)--\ref{M-fig3}(e) were computed from this simulation.

\item[Video S6]
Dragged-particle suspension simulation with $N = 128$.
The yellow arrows represent the unit orientation vectors of the particles.
The particle positions shown in the movies were recorded every 1000 time steps.
The angular pair distributions shown in Figs.~\ref{M-fig3}(f)--\ref{M-fig3}(h) were computed from this simulation.

\item[Video S7]
Simulation of an active-passive mixture, corresponding to the data used in Figs.~\ref{M-fig4}(e)~and~\ref{M-fig4}(f).
The red spheres represent active (dragged) particles, and the white spheres represent passive particles.
The particle positions shown in the movies were recorded every 1000 time steps.
The yellow arrows represent the unit orientation vectors of the particles.

\item[Video S8]
Neutral squirmer suspension simulation with $N = 128$.
The yellow arrows represent the unit orientation vectors of the particles.
The particle positions shown in the movies were recorded every 1000 time steps.
The time evolution of the polar order in Fig.~\ref{figS6}(b) and the angular pair distributions shown in Figs.~\ref{figS6}(d)--\ref{figS6}(f) were computed from this simulation.
\end{description}

\bibliography{ref}

%% file: Main_Text/Abstract.tex

\begin{abstract}
Spatial confinement profoundly impacts the transport and self-organization of active matter across diverse biological systems. While the collective orders in confined active matter have been extensively characterized, how geometric constraints reshape near-field flows and the resulting inter-particle correlations remains largely unexplored. In this study, we combine experiments and hydrodynamic simulations to investigate inter-particle correlations within quasi-two-dimensional \textit{Chlamydomonas reinhardtii} suspensions. We reveal two disentangled modes characterizing cell pairs: a dipolar mode and an entrainment mode, which exhibit a density- and distance-dependent competition. Combining single-cell flow field analysis, hydrodynamic simulations, and active-passive mixtures, we link these two modes to singular hydrodynamics and lubrication-induced entrainment. Our results demonstrate that spatiotemporal correlations in confined active matter are fundamentally rooted in the interplay of these two hydrodynamic mechanisms.
\end{abstract}

%% file: Main_Text/Introduction.tex

In natural environments, the transport and self-organization of swimming microorganisms typically manifest in confined spaces, such as those in surface-attached biofilms \cite{Hall-Stoodley2004,Mazza2016} and complex porous media \cite{Jin2024,Kumar2022}. The geometric constraints profoundly impact their motility and activate emergent orders by reshaping inter-particle interactions, especially hydrodynamic interactions \cite{Marchetti2013,Lauga2020,kim&karrila,Ishikawa2006}. In quasi-two-dimensional (quasi-2D) geometries, for instance, extensive studies have established how two parallel no-slip boundaries reconstruct the far-field flows of motile particles \cite{Liron1976,Cui2004,Brotto2013,Jeanneret2019} and give rise to a rich variety of collective orders \cite{Bricard2013,Shani2014,Nishiguchi2017,Baruah2026}. Concurrently, the scattering and trapping effects induced by static obstacles have also been widely characterized \cite{Takaha2023,Spagnolie2015}.

While various collective orders have been extensively characterized in confined active systems, how these population-level orders originate from individual-level interactions remains poorly understood. A primary reason this fundamental link remains elusive is that geometric constraints severely complicate the inter-particle interactions. Even for an individual microswimmer in confined spaces, developing an analytical understanding is heavily restricted by its complex force distribution \cite{Spagnolie2012,Brumley2014,Pradipta2026} and an infinite series of image singularities introduced by the boundaries \cite{Blake1971,Liron1976,Mathijssen2016,Takaha2023}. This limitation is drastically amplified when transitioning from a single-particle level to a many-body scale. In crowded environments, surrounding particles act as motile boundaries, deeply complicating the many-body hydrodynamics \cite{Shani2014,Yoshinaga2018,Kim2025} and making it exceedingly difficult to decipher how near-field flows drive particle correlations \cite{Kyoya2015,kim&karrila,Ishikawa2006}.

Here, combining experiments and hydrodynamic simulations on quasi-2D \textit{Chlamydomonas reinhardtii} suspensions [Figs.~\ref{fig1}(a)~and~\ref{fig1}(b)], we provide a pathway to decode many-body spatiotemporal correlations from a two-body perspective. For cell pairs, we identify two disentangled modes [Fig.~\ref{fig1}(c)~and~\ref{fig1}(d)]: a long-range dipolar mode (driven by 2D source dipole \cite{Cui2004,Liron1976,Brotto2013,Jeanneret2019}) and a short-range entrainment mode (by lubrication \cite{kim&karrila,Ishikawa2006}). Crucially, even within complex many-body dynamics, the signatures of both mechanisms remain preserved and disentangled in two-body correlations. By demonstrating how these multiscale mechanisms jointly shape inter-particle correlations, our work offers fundamental insights into spatiotemporal order in confined active systems.

%% file: Main_Text/Results.tex

\begin{figure*}[t!]
\centering
\includegraphics[width=\hsize]{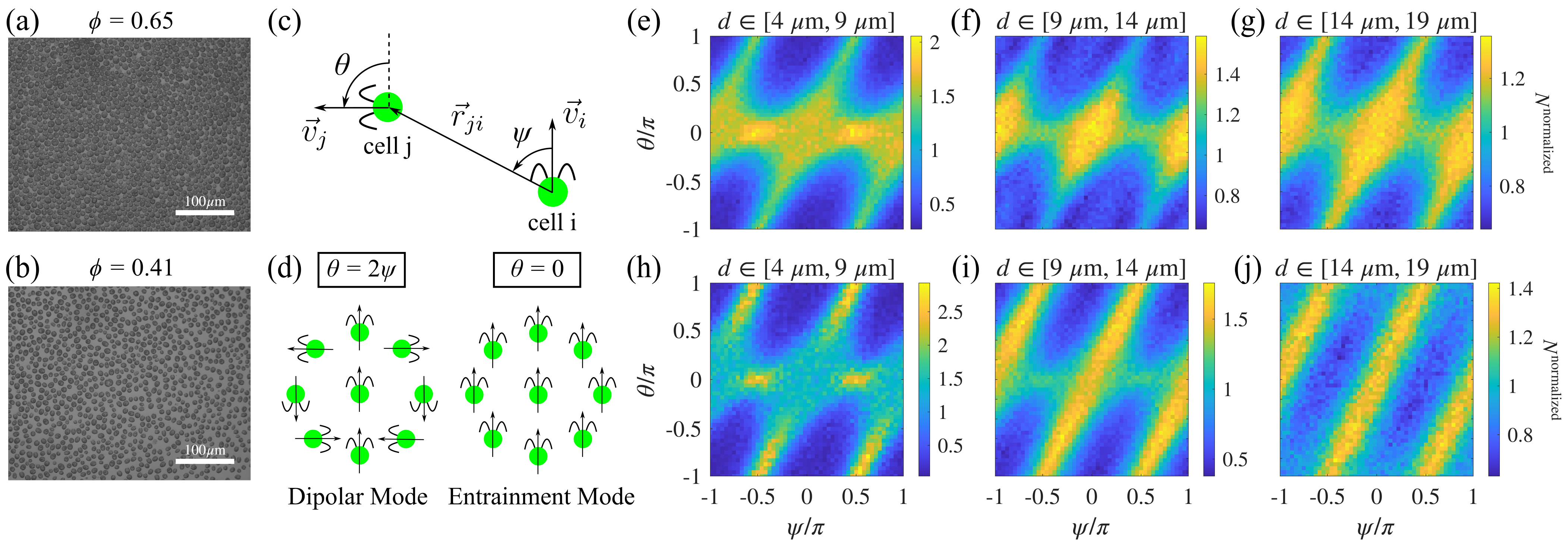}
\caption{Disentangled modes in quasi-2D \textit{C. reinhardtii} suspensions. (a),(b) Snapshots of quasi-2D \textit{C. reinhardtii} suspensions at area fraction (a) $\phi = 0.65$ and (b) $\phi = 0.41$. (c) Sketch of a cell pair ($i$, $j$) defining the oriented angles $\psi \in [-\pi, \pi]$ (from $\vec{v}_i$ to $\vec{r}_{ji}=\vec{r}_j-\vec{r}_i$) and $\theta \in [-\pi, \pi]$ (from $\vec{v}_i$ to $\vec{v}_j$). (d) Diagrams of the dipolar mode ($\theta = 2\psi$) and the entrainment mode ($\theta = 0$). (e)--(g) Angular pair distribution (area fraction $\phi = 0.65$) for distance ranges of (e) $d\in [4~\mathrm{\mu m}, 9~\mathrm{\mu m}]$, (f) $d\in [9~\mathrm{\mu m}, 14~\mathrm{\mu m}]$, and (g) $d \in [14~\mathrm{\mu m}, 19~\mathrm{\mu m}]$, where $N^{\mathrm{normalized}} = N(\psi,\theta)/\langle N(\psi,\theta)\rangle_{\psi, \theta}$ is the normalized pair counts per grid element. Here, $N(\psi,\theta)$ is the pair count within the given distance range $d$, and $\langle \cdots \rangle_{\psi, \theta}$ denotes the average over the entire angular domain. Distance ranges are determined based on the pair correlations [Figs.~\ref{S-figS1}(a)~and~\ref{S-figS1}(b)]. (h)--(j) Angular pair distribution (area fraction $\phi = 0.41$) for distance ranges of (h) $d\in [4~\mathrm{\mu m}, 9~\mathrm{\mu m}]$, (i) $d\in [9~\mathrm{\mu m}, 14~\mathrm{\mu m}]$, and (j) $d \in [14~\mathrm{\mu m}, 19~\mathrm{\mu m}]$.}
\label{fig1}
\end{figure*}

\textit{Experimental setup---}\textit{Chlamydomonas reinhardtii} cells (strain CC124, wild type, mean radius $\sim 5~\mathrm{\mu m}$) were confined in a quasi-2D chamber formed by two parallel PEG-coated glass slides separated by a $\sim 20~\mathrm{\mu m}$ gap. Concentrated suspensions was observed under a bright-field microscope with a red-light filter to prevent phototaxis. The cell suspension was mixed with $1$-$\mathrm{\mu m}$-diameter beads as flow tracers for particle image velocimetry (PIV) or $10$-$\mathrm{\mu m}$-diameter beads for the mixed systems to respectively investigate single-cell flow fields and active-passive mixtures. Complete details on experimental methods are provided in the Supplemental Material Sec.~A~and~B \cite{SM_ref}.

\textit{Two disentangled modes---}To characterize the spatiotemporal correlations within the confined active suspension, we jointly map the relative position angle $\psi$ and the relative velocity angle $\theta$ [Fig.~\ref{fig1}(c)] to obtain the angular pair distribution. For a given cell pair ($i$, $j$) with cell $i$ as the reference, these angles are kinematically defined using their center-of-mass velocities ($\vec{v}_i, \vec{v}_j$) instead of cell body orientations. As illustrated in Fig.~\ref{fig1}(c), $\psi$ corresponds to the angle from the reference cell's velocity $\vec{v}_i$ to the relative position vector $\vec{r}_{ji} = \vec{r}_j - \vec{r}_i$, while $\theta$ is the angle from $\vec{v}_i$ to $\vec{v}_j$. We find that the angular pair distribution is disentangled into two distinct modes [diagrams in Fig.~\ref{fig1}(d)]: an entrainment mode at $\theta = 0$, corresponding to a single horizontal characteristic line, and a dipolar mode at $\theta = 2\psi$, corresponding to three sloping characteristic lines defined by $\theta = 2\psi + 2n\pi \ (n = 0, \pm 1)$ within the angular domain [data in Figs.~\ref{fig1}(e)--\ref{fig1}(j)].

\begin{figure}[t!]
\centering
\includegraphics[width=\hsize]{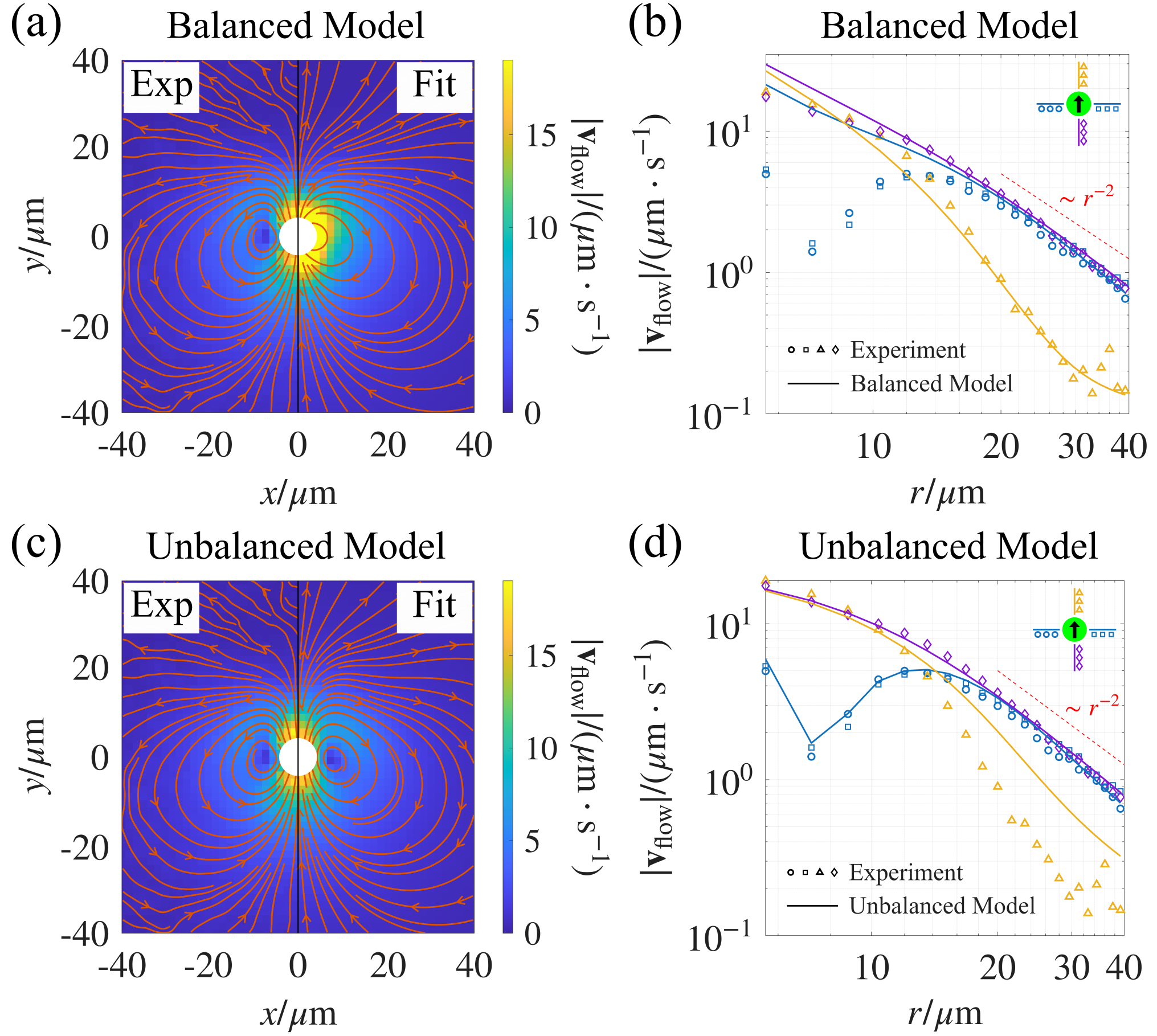}
\caption{Comparison between experimental and fitting flows around a single \textit{C. reinhardtii} cell. Fitting procedures and results are in Sec.~E and Table~\ref{S-tab:fit_results} of Supplemental Material \cite{SM_ref}. (a) Left panel: the experimental flow; Right panel: the fitting flow of the balanced model including a source-dipole and three regularized Brinkmanlets ($F = 2F^{\prime}$) (fitting region: $r\in [20~\mathrm{\mu m}, 40~\mathrm{\mu m}]$). All panels show velocity magnitude (background color) and streamlines (red curves) of the flow field. The $x, y$ components are defined in the local body coordinate system, with the origin at the body's center of mass and $y$-axis along the direction of the cell's smoothed velocity (smoothing window size: $0.1~\mathrm{s}$). (b) Experimental measurements (symbols) and balanced model fits (lines) of flow profiles along four directions extracted from (a), indicated by separate colors and symbols in the inset. (c) Left panel: the experimental flow; Right panel: the fitting flow of the unbalanced model including a source-dipole and three regularized Brinkmanlets ($F> 2F^{\prime}$) (fitting region: $r\in [5~\mathrm{\mu m}, 40~\mathrm{\mu m}]$). Diagram settings follow (a). (d) Experimental measurements (symbols) and unbalanced model fits (lines) of flow profiles along four directions extracted from (c), indicated by separate colors and symbols in the inset.}
\label{fig2}
\end{figure}

These two modes exhibit a competition governed by both the global area fraction $\phi$ and the local distance $d$ between centers of mass of cell pairs. At a high area fraction ($\phi = 0.65$) with an isotropic pair correlation [Fig.~\ref{S-figS1}(a) in Supplemental Material \cite{SM_ref}], the short-range pair distribution ($d \in [4~\mathrm{\mu m}, 9~\mathrm{\mu m}]$) is dominated by the $\theta = 0$ mode [\figpref{fig1}{e}], indicating alignments between adjacent microswimmers. This pair distribution decays rapidly as the distance increases to longer ranges ($d \in [9~\mathrm{\mu m}, 14~\mathrm{\mu m}]$ and $d \in [14~\mathrm{\mu m}, 19~\mathrm{\mu m}]$) [Figs.~\ref{fig1}(f)~and~\ref{fig1}(g)]. Conversely, at a lower area fraction ($\phi = 0.41$) with an anisotropic pair correlation [Fig.~\ref{S-figS1}(b)], the near-field $\theta = 0$ signature weakens [\figpref{fig1}{h}], and the $\theta = 2\psi$ mode emerges prominently at larger distances [Figs.~\ref{fig1}(i)~and~\ref{fig1}(j)]. This density- and distance-dependent competition is quantitatively characterized by the velocity correlation and the normalized lateral pair correlation for two modes (see Sec.~C and Fig.~\ref{S-figS1} of Supplemental Material \cite{SM_ref}), revealing a crossover from a ($\theta = 2\psi$)-dominated to a ($\theta = 0$)-dominated regime with increasing density.

\textit{Flow field analysis---}To establish the hydrodynamic origins of these two modes, we map the flow field around a single \textit{C. reinhardtii} cell using the high-speed recording ($500~\mathrm{fps}$) and PIV. Under quasi-2D confinements, the synchronous beating of the dual flagella of our \textit{C. reinhardtii} cell is disrupted, leading to a swimming state of ``wobbler'' \cite{Mondal2021}. This state features asynchronous flagellar beating and body wobbling, manifested as zigzag trajectories and periodic angular oscillations between the cell orientation and smoothed velocity [Fig.~\ref{S-figS3_oldS2}(a)~and~Video~S3]. By defining the phase $\varphi_x$ based on the lateral oscillation [Fig.~\ref{S-figS3_oldS2}(b)], we calculate both phase-resolved time-averaged flow fields [Figs.~\ref{S-figS3_oldS2}(c)--\ref{S-figS3_oldS2}(f)] and an all-frame time-averaged flow field [Figs.~\ref{fig2}(a),~\ref{fig2}(c)~and~\ref{S-figS4_oldS3}(a)]. The time-averaged field exhibits distinct spatial regimes: the far-field flow displays a $\theta = 2\psi$ signature with $\sim r^{-2}$ power-law decay [Figs.~\ref{fig2}(b)~and~\ref{fig2}(d)] \cite{footnote_flow_powerlaw}, while the near-field flow shows a front-back asymmetry, featuring strong longitudinal flows aligned with the swimming direction and two distinct lateral vortices [Figs.~\ref{fig2}(a),~\ref{fig2}(c)~and~\ref{S-figS4_oldS3}(a)]. The $\theta = 2\psi$ signature of single-cell flows hints that the dipolar mode in the angular correlation is strongly linked to singular hydrodynamics, primarily driven by the 2D source dipole that inherently carries the $\theta=2\psi$ signature \cite{Cui2004,Liron1976,Brotto2013,Jeanneret2019}.

To fit this time-averaged flow field, we utilize three regularized Brinkmanlets and a 2D source dipole \cite{SM_ref,Leiderman2016}. The Brinkmanlets serve as fundamental force solutions that incorporate the top and bottom no-slip boundaries via a 2D depth-average \cite{Jeanneret2019,Nagel2015,Leiderman2016}. Meanwhile, the source dipole is required to fulfill mass conservation, as the finite-size cell body moves at a velocity different from the background fluid \cite{Cui2004,Liron1976,Brotto2013,Jeanneret2019}. We assume a configuration with one force $F$ at the center of mass pointing along the velocity, and two forces $F^{\prime}$ at the flagella pointing oppositely. The physically consistent balanced model ($F = 2F^{\prime}$) \cite{Drescher2010,Jeanneret2019,Pradipta2026} accurately captures the far-field flow (fitting range: $r \in [20~\mathrm{\mu m}, 40~\mathrm{\mu m}]$) [Figs.~\ref{fig2}(a)~and~\ref{fig2}(b)], but yields significant deviations in the near-field flow. Extending the fitting range to the near-field region does not help, as the fitting result then fails to describe the far-field flow [Fig.~\ref{S-figS}]. To test if the near-field flow can be resolved, we also evaluate an unbalanced model ($F > 2F^{\prime}$) as an approximation. This alternative successfully fits the entire spatial range (fitting range: $r \in [5~\mathrm{\mu m}, 40~\mathrm{\mu m}]$), reproducing the near-field front-back asymmetry, lateral vortices [\figpref{fig2}{c}], and velocity magnitudes along all four directions down to $r = 5~\mathrm{\mu m}$ [\figpref{fig2}{d}] \cite{footnote_flow_powerlaw}.

The empirical effectiveness of the unbalanced model can be qualitatively understood from both near- and far-field perspectives. In the near field, the flagellar forcing is widely distributed, especially for ``wobblers'' \cite{Mondal2021}, which allows the localized cell-body force to dominate the near-field flow. Furthermore, in the far field, the lack of exact force balance is unharmful. The net unbalanced force $F-2F^{\prime}>0$ effectively acts as a 2D source dipole, since both Brinkmanlet and source dipole decay into 2D irrotational potential flows with the identical Oseen tensor $(\mathbbm{1}-2\hat{\boldsymbol{r}}\hat{\boldsymbol{r}})/r^2$ \cite{Liron1976,SM_ref}, naturally preserving the $\theta = 2\psi$ asymptotic signature at large distances.

\begin{figure*}[t!]
\centering
\includegraphics[width=0.85\hsize]{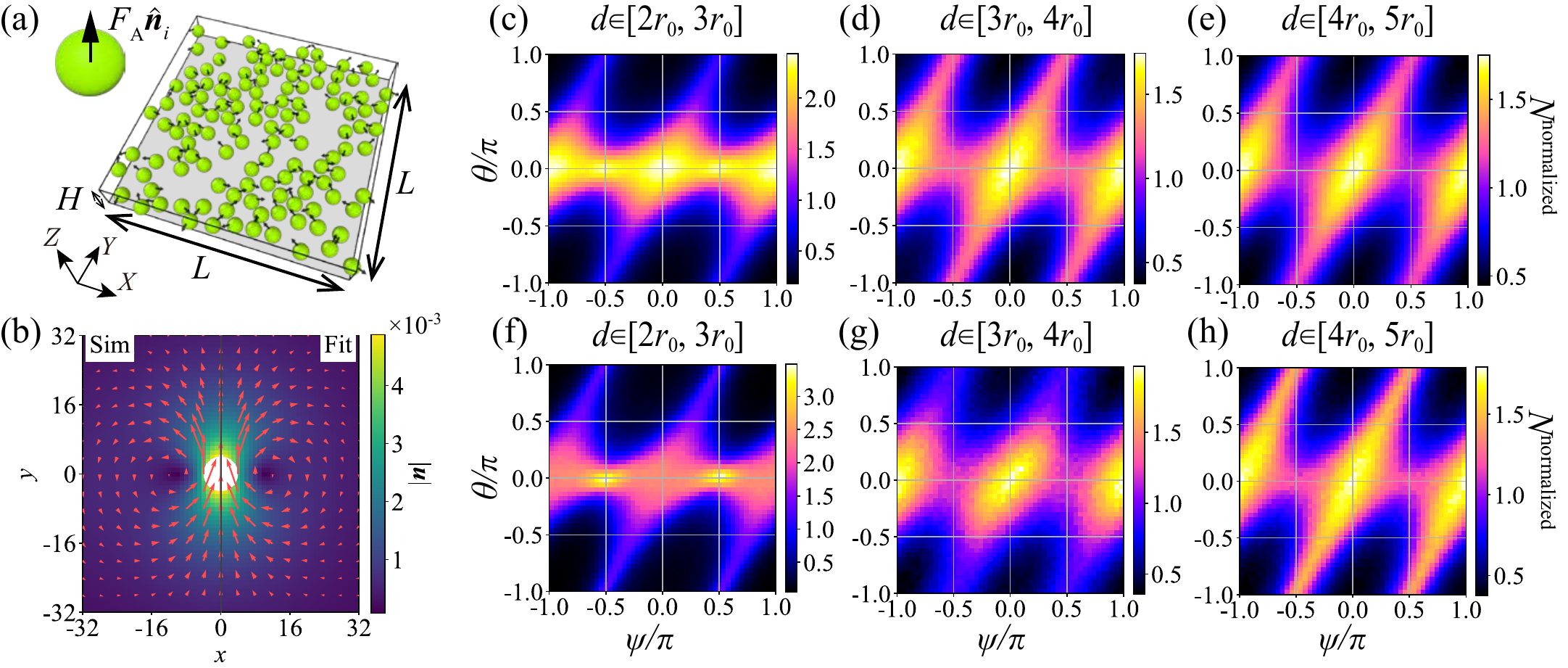}
\caption{Simulation setup for dragged-particles suspension and the results. (a) Simulation system composed of $N$ dragged-particle in quasi-2D system ($L\times L \times H = 128 \times 128 \times 15$). (b) Left panel: the depth-averaged flow field around a single dragged-particle at $z=0$, obtained from a simulation in the absence of rotational diffusion. Right panel: best-fit flow field generated by a combination of a single regularized Brinkmanlet and a source dipole at the particle center \cite{SM_ref,Leiderman2016}. (c)-(e) Angular pair distribution of $N=224\,(\phi=0.69)$ dragged-particles suspension at the ranges of (c) $d \in [2r_0,3r_0]$, (d) $d \in [3r_0,4r_0]$, and (e) $d \in [4r_0,5r_0]$. (f)-(h) Angular pair distribution of $N=128\,(\phi=0.39)$ dragged-particles suspension at the ranges of (f) $d \in [2r_0,3r_0]$, (g) $d \in [3r_0,4r_0]$, and (h) $d \in [4r_0,5r_0]$.}
\label{fig3}
\end{figure*}
\textit{Simulation---}To further investigate the origins of the two modes, we perform hydrodynamic simulations using a model microswimmer system. Based on our experimental observation that the flow field around a single cell was better approximated by an unbalanced-force model [Figs.~\ref{fig2}(c)~and~\ref{fig2}(d)], we adopt a dragged-particle model [\figpref{fig3}{a}]. In this model, spherical particles of radius $r_0$ are subjected to a constant active force, $F_{\rm A} \hat{\boldsymbol{n}}$. Here, $\hat{\boldsymbol{n}}$ denotes the unit orientation vector and $F_{\rm A}$ is the magnitude of the active force. The orientation vectors are confined to the $XY$ plane so that the particle orientations remain parallel to the confining walls, and are subject to rotational diffusion to suppress polar ordering. The particles interact via a soft-core potential to prohibit overlapping between particles:
$U_{\rm sc}(r_{ij}) = \epsilon_0 (\sigma/r_{ij})^{12}$, where $\sigma=2r_0$ denotes the diameter of particles and $r_{ij}$ is the distance between $i$th and $j$th spheres. The particles also interact through the host fluid via hydrodynamic interactions. In our simulations, these hydrodynamic interactions are incorporated using the smoothed profile method \cite{Nakayama2005, Yamamoto2021}, which directly solves the Navier-Stokes equation. Note that, in this model, the particles interact only through short-range repulsive interactions and hydrodynamic interactions, with no steric torque.

As shown in \figpref{fig3}{a}, our simulation system is confined between two stationary planar walls located at $Z=\pm H/2$. No-slip boundary conditions are imposed on the walls: $\boldsymbol{u}(X,Y,\pm H/2) = \boldsymbol{0}$. Periodic boundary conditions are imposed in the $X$ and $Y$ directions, with lateral dimensions $L\times L$.

The particles and the boundary walls interact via the same soft-core potential as that of inter-particles, with $\sigma=r_0+d_0$. Here, $d_0$ is a buffer length to maintain a finite gap between the particle surfaces and the walls; consequently, particle centers are restricted to $[-H/2+(d_0+r_0), H/2-(d_0+r_0)]$. We set $r_0=4$, $d_0=2$, $F_{\rm A}=1$, $\epsilon_0=2$, $L=128$, and $H=15$, with the remaining parameters specified in Sec.~F of Supplemental Material \cite{SM_ref}. The resulting Reynolds number, $Re \simeq 0.08$, is sufficiently small.

We first examined the flow field around a single particle, as shown in \figpref{fig3}{b}. The depth-averaged flow field exhibits two vortices at $\sim r_0$, similar to those observed in our experiments [Figs.~\ref{fig2}(a)~and~\ref{fig2}(c)]. We fit the flow field using a single regularized Brinkmanlet at the center of a sphere, and it shows close agreement with the numerical result [\figpref{fig3}{b}] (Fitting procedures and results are shown in Sec.~E of Supplemental Material \cite{SM_ref}).

Figures \ref{fig3}(c)-\ref{fig3}(e) show the angular pair distribution for a suspension of $N=224$ dragged particles $(\phi=0.69)$. The numerical results are consistent with our experimental observations: at the short range ($d<3r_0$), the $\theta=0$ mode is dominant [\figpref{fig3}{c}], and this mode sharply decays with increasing $d$ [Fig.~\ref{fig3}(d)]. In contrast, the $\theta=2\psi$ mode is enhanced as $d$ increases [\figpref{fig3}{e}]. Regarding the density dependence, the $\theta=0$ mode weakens at near-field [Fig.~\ref{fig3}(f)] while the $\theta=2\psi$ mode is enhanced at far-field [Fig.~\ref{fig3}(h)] for a lower area fraction ($N=128, \phi=0.39$). Our simulations revealed that the spherical dragged model, despite being a highly simplified representation of \textit{C. reinhardtii}, exhibits the same two modes observed experimentally. We therefore conclude that these modes arise purely from hydrodynamic interactions between spherical bodies. Near-field hydrodynamic interactions give rise to the entrainment mode ($\theta = 0$), whereas far-field hydrodynamic interactions generated by an isolated particle account for the dipolar mode ($\theta = 2\psi$).

\begin{figure}[t!]
\centering
\includegraphics[width=\hsize]{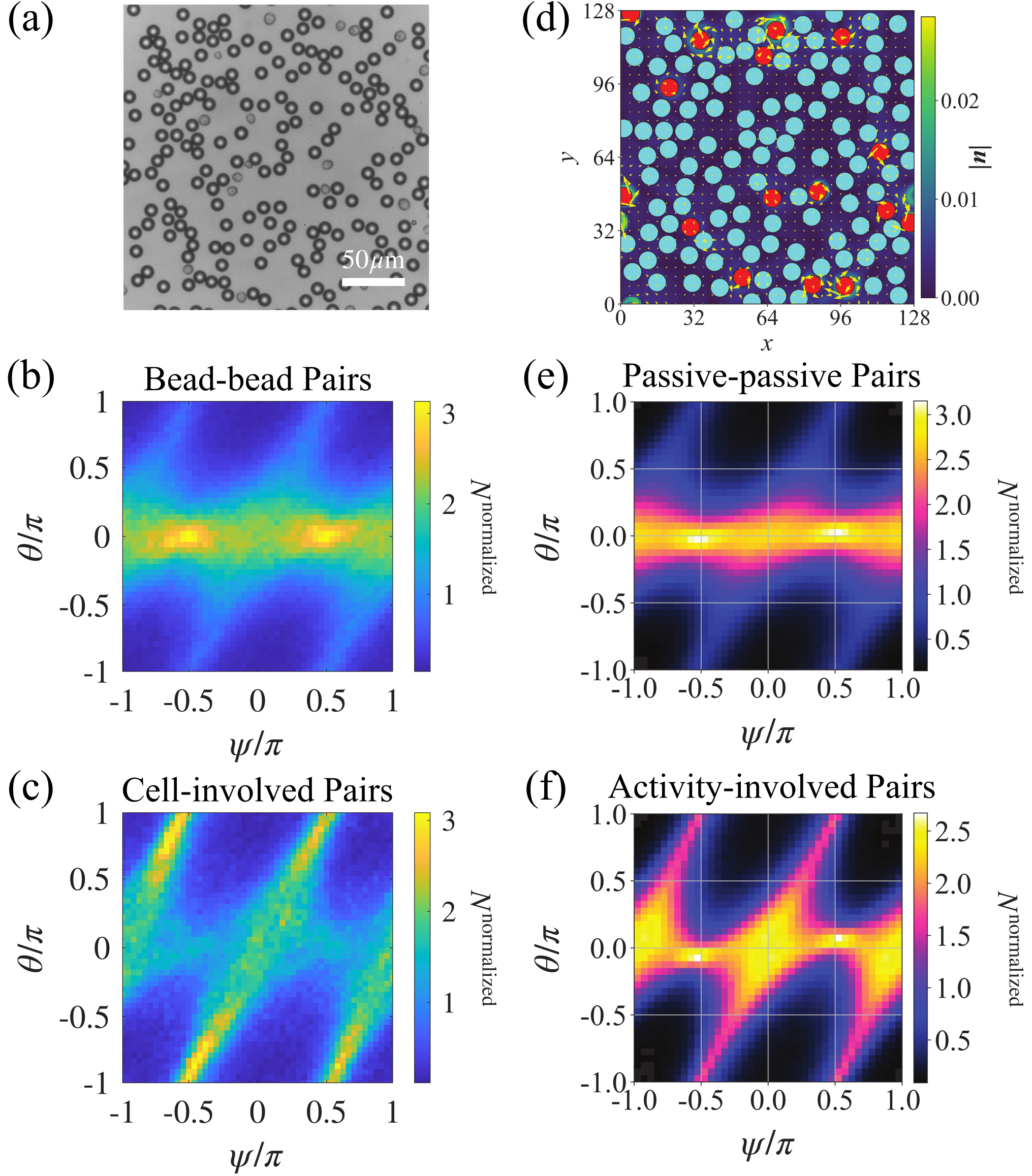}
\caption{Separation of two modes in an active-passive mixture. (a) Experimental snapshot of quasi-2D mixtures combining active \textit{C. reinhardtii} cells (mean number fraction of $12.5\%$) and $10$-$\mathrm{\mu m}$-diameter passive beads. (b) Angular pair distribution of bead-bead pairs ($d\in[4~\mathrm{\mu m}, 14~\mathrm{\mu m}]$), showing a strong $\theta = 0$ mode and a weak $\theta = 2\psi$ mode. (c) Angular pair distribution of cell-involved pairs ($d\in[4~\mathrm{\mu m}, 14~\mathrm{\mu m}]$), highlighting the emergence of the $\theta = 2\psi$ mode. (d) Simulation snapshot of quasi-2D mixtures combining active dragged particle (red circles) and passive particle (blue circles). For the flow field, the yellow arrows show the local velocity vectors and the background color corresponds to the magnitude $\vert \boldsymbol{u}\vert$. (e) Angular pair distribution of passive-passive pairs ($d\in [2r_0,4r_0]$) in the simulation, also showing a strong $\theta = 0$ mode and a weak $\theta = 2\psi$ mode. (f) Angular pair distribution of activity-involved pairs ($d\in [2r_0,4r_0]$) in the simulation, also showing the emergence of the $\theta = 2\psi$ mode.}
\label{fig4}
\end{figure}

\textit{Active-passive mixture---}To investigate physical mechanisms underlying these two modes, we introduce a quasi-2D mixture comprising a minority of active cells (mean number fraction of $12.5\%$) and a majority of passive beads [\figpref{fig4}{a}]. Experimentally, they can be clearly distinguished based on the radial profile of the intensity. We subtract the background flow ($\sim 3~\mathrm{\mu m/s}$, evaluated from ensemble-averaged passive-bead velocities) to obtain trajectories of active cells (mean speed $\sim 25~\mathrm{\mu m/s}$) and passive beads ($\sim 4~\mathrm{\mu m/s}$).

Evaluating the angular pair distributions for distinct sub-populations turns out to disentangle the two modes. For bead-bead pairs, the angular pair distribution is heavily dominated by the $\theta = 0$ mode, with the $\theta = 2\psi$ signature essentially absent [\figpref{fig4}{b}]. In stark contrast, when evaluating pairs that include active cells, the $\theta = 2\psi$ mode robustly emerges [\figpref{fig4}{c}]. These results demonstrate that activity is essential to produce the $\theta = 2\psi$ mode, while the $\theta = 0$ mode is relatively activity-independent. 

We numerically confirm that these behaviors can also be observed in a mixture of dragged and passive particles. We simulate a system of $N=128$ particles, including $N_{\rm a}=16$ active (dragged) particles, under the same conditions as those shown in \figpref{fig3}{a}. For passive-passive pairs, the $\theta=0$ mode is pronounced, whereas the $\theta=2\psi$ mode is weak [\figpref{fig4}{e}]. By contrast, the $\theta=2\psi$ mode emerges for activity-involved pairs [\figpref{fig4}{f}].

Crucially, our simulation defines the steric interactions between spherical particles as purely radial repulsions, which cannot contribute to the observed alignment. The effective alignment must therefore be hydrodynamically mediated. Given its near-field localization, we attribute the $\theta = 0$ mode to lubrication-induced entrainment: the fluid confined between close surfaces drives a hydrodynamic entrainment that yields this alignment (see Sec.~G and Fig.~\ref{S-figS5} of Supplemental Material \cite{SM_ref}).

%% file: Main_Text/Summary-Discussion.tex
\textit{Discussion---}Based on the statistical analysis of two-body angular correlations, we identify two disentangled modes in quasi-2D \textit{C. reinhardtii} suspensions: a dipolar mode ($\theta = 2\psi$) and an entrainment mode ($\theta = 0$) [Fig.~\ref{fig1}(d)]. Under quasi-2D confinements, the single-cell flow field exhibits a characteristic $\theta = 2\psi$ signature in the far field (Figs.~\ref{fig2} and \ref{S-figS4_oldS3}), which is imprinted onto angular correlations of cell pairs. By combining hydrodynamic simulations (Fig.~\ref{fig3}) and active-passive mixtures (Fig.~\ref{fig4}), we demonstrate that the effective alignment ($\theta = 0$) is a generic property originating from hydrodynamics at the near-field limit. We argue that the $\theta = 0$ mode originates from lubrication effects of the fluid between two nearby particles (see Sec.~G and Fig.~\ref{S-figS5} of Supplemental Material \cite{SM_ref}). Our work elucidates that the signatures of hydrodynamic interactions could be mapped onto and dictate spatiotemporal correlations.

Despite being a many-body suspension dominated by near-field interactions, our system strongly exhibits the $\theta = 2\psi$ mode in angular pair correlation, which is a signature of a 2D source dipole rooted in far-field singular hydrodynamics \cite{Cui2004,Liron1976,Brotto2013,Jeanneret2019}. While analogous signatures also emerge in dragged 2D water-in-oil droplets explained by 2D dipole potential \cite{Shani2014}, how boundary conditions of surrounding particles modulate these many-body interactions remains poorly understood. Furthermore, although our simulations show that lubrication entrainment suffices to trigger the effective alignment ($\theta = 0$), real cellular interactions are likely more intricate. In high-density regimes, steric shearing torques \cite{Dauchot2005} and steric-induced spontaneous alignment \cite{Caprini2020} may also contribute to the $\theta = 0$ mode alongside hydrodynamic lubrication. 

Ultimately, unraveling how many-body effects \cite{Shani2014,Yoshinaga2018,Kim2025} shape collective dynamics remains a promising frontier. By bridging individual interactions and spatiotemporal correlations from a two-body perspective, our work establishes a powerful framework for deciphering complex many-body dynamics in active matter.

%% file: main-arxiv.bbl
\begin{thebibliography}{36}%
\makeatletter
\providecommand \@ifxundefined [1]{%
 \@ifx{#1\undefined}
}%
\providecommand \@ifnum [1]{%
 \ifnum #1\expandafter \@firstoftwo
 \else \expandafter \@secondoftwo
 \fi
}%
\providecommand \@ifx [1]{%
 \ifx #1\expandafter \@firstoftwo
 \else \expandafter \@secondoftwo
 \fi
}%
\providecommand \natexlab [1]{#1}%
\providecommand \enquote  [1]{``#1''}%
\providecommand \bibnamefont  [1]{#1}%
\providecommand \bibfnamefont [1]{#1}%
\providecommand \citenamefont [1]{#1}%
\providecommand \href@noop [0]{\@secondoftwo}%
\providecommand \href [0]{\begingroup \@sanitize@url \@href}%
\providecommand \@href[1]{\@@startlink{#1}\@@href}%
\providecommand \@@href[1]{\endgroup#1\@@endlink}%
\providecommand \@sanitize@url [0]{\catcode `\\12\catcode `\$12\catcode `\&12\catcode `\#12\catcode `\^12\catcode `\_12\catcode `\%12\relax}%
\providecommand \@@startlink[1]{}%
\providecommand \@@endlink[0]{}%
\providecommand \url  [0]{\begingroup\@sanitize@url \@url }%
\providecommand \@url [1]{\endgroup\@href {#1}{\urlprefix }}%
\providecommand \urlprefix  [0]{URL }%
\providecommand \Eprint [0]{\href }%
\providecommand \doibase [0]{https://doi.org/}%
\providecommand \selectlanguage [0]{\@gobble}%
\providecommand \bibinfo  [0]{\@secondoftwo}%
\providecommand \bibfield  [0]{\@secondoftwo}%
\providecommand \translation [1]{[#1]}%
\providecommand \BibitemOpen [0]{}%
\providecommand \bibitemStop [0]{}%
\providecommand \bibitemNoStop [0]{.\EOS\space}%
\providecommand \EOS [0]{\spacefactor3000\relax}%
\providecommand \BibitemShut  [1]{\csname bibitem#1\endcsname}%
\let\auto@bib@innerbib\@empty
\bibitem [{\citenamefont {Hall-Stoodley}\ \emph {et~al.}(2004)\citenamefont {Hall-Stoodley}, \citenamefont {Costerton},\ and\ \citenamefont {Stoodley}}]{Hall-Stoodley2004}%
  \BibitemOpen
  \bibfield  {author} {\bibinfo {author} {\bibfnamefont {L.}~\bibnamefont {Hall-Stoodley}}, \bibinfo {author} {\bibfnamefont {J.~W.}\ \bibnamefont {Costerton}},\ and\ \bibinfo {author} {\bibfnamefont {P.}~\bibnamefont {Stoodley}},\ }\bibfield  {title} {\bibinfo {title} {Bacterial biofilms: from the natural environment to infectious diseases},\ }\href {https://doi.org/10.1038/nrmicro821} {\bibfield  {journal} {\bibinfo  {journal} {Nat. Rev. Microbiol.}\ }\textbf {\bibinfo {volume} {2}},\ \bibinfo {pages} {95} (\bibinfo {year} {2004})}\BibitemShut {NoStop}%
\bibitem [{\citenamefont {Mazza}(2016)}]{Mazza2016}%
  \BibitemOpen
  \bibfield  {author} {\bibinfo {author} {\bibfnamefont {M.~G.}\ \bibnamefont {Mazza}},\ }\bibfield  {title} {\bibinfo {title} {The physics of biofilms—an introduction},\ }\href {https://doi.org/10.1088/0022-3727/49/20/203001} {\bibfield  {journal} {\bibinfo  {journal} {J. Phys. D: Appl. Phys.}\ }\textbf {\bibinfo {volume} {49}},\ \bibinfo {pages} {203001} (\bibinfo {year} {2016})}\BibitemShut {NoStop}%
\bibitem [{\citenamefont {Jin}\ and\ \citenamefont {Sengupta}(2024)}]{Jin2024}%
  \BibitemOpen
  \bibfield  {author} {\bibinfo {author} {\bibfnamefont {C.}~\bibnamefont {Jin}}\ and\ \bibinfo {author} {\bibfnamefont {A.}~\bibnamefont {Sengupta}},\ }\bibfield  {title} {\bibinfo {title} {Microbes in porous environments: from active interactions to emergent feedback},\ }\href {https://doi.org/10.1007/s12551-024-01185-7} {\bibfield  {journal} {\bibinfo  {journal} {Biophys. Rev.}\ }\textbf {\bibinfo {volume} {16}},\ \bibinfo {pages} {173} (\bibinfo {year} {2024})}\BibitemShut {NoStop}%
\bibitem [{\citenamefont {Kumar}\ \emph {et~al.}(2022)\citenamefont {Kumar}, \citenamefont {Guasto},\ and\ \citenamefont {Ardekani}}]{Kumar2022}%
  \BibitemOpen
  \bibfield  {author} {\bibinfo {author} {\bibfnamefont {M.}~\bibnamefont {Kumar}}, \bibinfo {author} {\bibfnamefont {J.~S.}\ \bibnamefont {Guasto}},\ and\ \bibinfo {author} {\bibfnamefont {A.~M.}\ \bibnamefont {Ardekani}},\ }\bibfield  {title} {\bibinfo {title} {Transport of complex and active fluids in porous media},\ }\href {https://doi.org/10.1122/8.0000389} {\bibfield  {journal} {\bibinfo  {journal} {J. Rheol.}\ }\textbf {\bibinfo {volume} {66}},\ \bibinfo {pages} {375} (\bibinfo {year} {2022})}\BibitemShut {NoStop}%
\bibitem [{\citenamefont {Marchetti}\ \emph {et~al.}(2013)\citenamefont {Marchetti}, \citenamefont {Joanny}, \citenamefont {Ramaswamy}, \citenamefont {Liverpool}, \citenamefont {Prost}, \citenamefont {Rao},\ and\ \citenamefont {Simha}}]{Marchetti2013}%
  \BibitemOpen
  \bibfield  {author} {\bibinfo {author} {\bibfnamefont {M.~C.}\ \bibnamefont {Marchetti}}, \bibinfo {author} {\bibfnamefont {J.~F.}\ \bibnamefont {Joanny}}, \bibinfo {author} {\bibfnamefont {S.}~\bibnamefont {Ramaswamy}}, \bibinfo {author} {\bibfnamefont {T.~B.}\ \bibnamefont {Liverpool}}, \bibinfo {author} {\bibfnamefont {J.}~\bibnamefont {Prost}}, \bibinfo {author} {\bibfnamefont {M.}~\bibnamefont {Rao}},\ and\ \bibinfo {author} {\bibfnamefont {R.~A.}\ \bibnamefont {Simha}},\ }\bibfield  {title} {\bibinfo {title} {Hydrodynamics of soft active matter},\ }\href {https://doi.org/10.1103/RevModPhys.85.1143} {\bibfield  {journal} {\bibinfo  {journal} {Rev. Mod. Phys.}\ }\textbf {\bibinfo {volume} {85}},\ \bibinfo {pages} {1143} (\bibinfo {year} {2013})}\BibitemShut {NoStop}%
\bibitem [{\citenamefont {Lauga}(2020)}]{Lauga2020}%
  \BibitemOpen
  \bibfield  {author} {\bibinfo {author} {\bibfnamefont {E.}~\bibnamefont {Lauga}},\ }\href@noop {} {\emph {\bibinfo {title} {The Fluid Dynamics of Cell Motility}}},\ Cambridge Texts in Applied Mathematics\ (\bibinfo  {publisher} {Cambridge University Press},\ \bibinfo {year} {2020})\BibitemShut {NoStop}%
\bibitem [{\citenamefont {Kim}\ and\ \citenamefont {Karrila}(1991)}]{kim&karrila}%
  \BibitemOpen
  \bibfield  {author} {\bibinfo {author} {\bibfnamefont {S.}~\bibnamefont {Kim}}\ and\ \bibinfo {author} {\bibfnamefont {S.}~\bibnamefont {Karrila}},\ }\href@noop {} {\emph {\bibinfo {title} {Microhydrodynamics: Principles and Selected Applications}}},\ Butterworth-Heinemann series in chemical engineering\ (\bibinfo  {publisher} {Elsevier Science \& Technology Books},\ \bibinfo {year} {1991})\BibitemShut {NoStop}%
\bibitem [{\citenamefont {Ishikawa}\ \emph {et~al.}(2006)\citenamefont {Ishikawa}, \citenamefont {Simmonds},\ and\ \citenamefont {Pedley}}]{Ishikawa2006}%
  \BibitemOpen
  \bibfield  {author} {\bibinfo {author} {\bibfnamefont {T.}~\bibnamefont {Ishikawa}}, \bibinfo {author} {\bibfnamefont {M.~P.}\ \bibnamefont {Simmonds}},\ and\ \bibinfo {author} {\bibfnamefont {T.~J.}\ \bibnamefont {Pedley}},\ }\bibfield  {title} {\bibinfo {title} {Hydrodynamic interaction of two swimming model micro-organisms},\ }\href {https://doi.org/10.1017/S0022112006002631} {\bibfield  {journal} {\bibinfo  {journal} {J. Fluid Mech.}\ }\textbf {\bibinfo {volume} {568}},\ \bibinfo {pages} {119–160} (\bibinfo {year} {2006})}\BibitemShut {NoStop}%
\bibitem [{\citenamefont {Liron}\ and\ \citenamefont {Mochon}(1976)}]{Liron1976}%
  \BibitemOpen
  \bibfield  {author} {\bibinfo {author} {\bibfnamefont {N.}~\bibnamefont {Liron}}\ and\ \bibinfo {author} {\bibfnamefont {S.}~\bibnamefont {Mochon}},\ }\bibfield  {title} {\bibinfo {title} {Stokes flow for a stokeslet between two parallel flat plates},\ }\href {https://doi.org/10.1007/BF01535565} {\bibfield  {journal} {\bibinfo  {journal} {J. Eng. Math.}\ }\textbf {\bibinfo {volume} {10}},\ \bibinfo {pages} {287} (\bibinfo {year} {1976})}\BibitemShut {NoStop}%
\bibitem [{\citenamefont {Cui}\ \emph {et~al.}(2004)\citenamefont {Cui}, \citenamefont {Diamant}, \citenamefont {Lin},\ and\ \citenamefont {Rice}}]{Cui2004}%
  \BibitemOpen
  \bibfield  {author} {\bibinfo {author} {\bibfnamefont {B.}~\bibnamefont {Cui}}, \bibinfo {author} {\bibfnamefont {H.}~\bibnamefont {Diamant}}, \bibinfo {author} {\bibfnamefont {B.}~\bibnamefont {Lin}},\ and\ \bibinfo {author} {\bibfnamefont {S.~A.}\ \bibnamefont {Rice}},\ }\bibfield  {title} {\bibinfo {title} {Anomalous hydrodynamic interaction in a quasi-two-dimensional suspension},\ }\href {https://doi.org/10.1103/PhysRevLett.92.258301} {\bibfield  {journal} {\bibinfo  {journal} {Phys. Rev. Lett.}\ }\textbf {\bibinfo {volume} {92}},\ \bibinfo {pages} {258301} (\bibinfo {year} {2004})}\BibitemShut {NoStop}%
\bibitem [{\citenamefont {Brotto}\ \emph {et~al.}(2013)\citenamefont {Brotto}, \citenamefont {Caussin}, \citenamefont {Lauga},\ and\ \citenamefont {Bartolo}}]{Brotto2013}%
  \BibitemOpen
  \bibfield  {author} {\bibinfo {author} {\bibfnamefont {T.}~\bibnamefont {Brotto}}, \bibinfo {author} {\bibfnamefont {J.-B.}\ \bibnamefont {Caussin}}, \bibinfo {author} {\bibfnamefont {E.}~\bibnamefont {Lauga}},\ and\ \bibinfo {author} {\bibfnamefont {D.}~\bibnamefont {Bartolo}},\ }\bibfield  {title} {\bibinfo {title} {Hydrodynamics of confined active fluids},\ }\href {https://doi.org/10.1103/PhysRevLett.110.038101} {\bibfield  {journal} {\bibinfo  {journal} {Phys. Rev. Lett.}\ }\textbf {\bibinfo {volume} {110}},\ \bibinfo {pages} {038101} (\bibinfo {year} {2013})}\BibitemShut {NoStop}%
\bibitem [{\citenamefont {Jeanneret}\ \emph {et~al.}(2019)\citenamefont {Jeanneret}, \citenamefont {Pushkin},\ and\ \citenamefont {Polin}}]{Jeanneret2019}%
  \BibitemOpen
  \bibfield  {author} {\bibinfo {author} {\bibfnamefont {R.}~\bibnamefont {Jeanneret}}, \bibinfo {author} {\bibfnamefont {D.~O.}\ \bibnamefont {Pushkin}},\ and\ \bibinfo {author} {\bibfnamefont {M.}~\bibnamefont {Polin}},\ }\bibfield  {title} {\bibinfo {title} {Confinement enhances the diversity of microbial flow fields},\ }\href {https://doi.org/10.1103/PhysRevLett.123.248102} {\bibfield  {journal} {\bibinfo  {journal} {Phys. Rev. Lett.}\ }\textbf {\bibinfo {volume} {123}},\ \bibinfo {pages} {248102} (\bibinfo {year} {2019})}\BibitemShut {NoStop}%
\bibitem [{\citenamefont {Bricard}\ \emph {et~al.}(2013)\citenamefont {Bricard}, \citenamefont {Caussin}, \citenamefont {Desreumaux}, \citenamefont {Dauchot},\ and\ \citenamefont {Bartolo}}]{Bricard2013}%
  \BibitemOpen
  \bibfield  {author} {\bibinfo {author} {\bibfnamefont {A.}~\bibnamefont {Bricard}}, \bibinfo {author} {\bibfnamefont {J.-B.}\ \bibnamefont {Caussin}}, \bibinfo {author} {\bibfnamefont {N.}~\bibnamefont {Desreumaux}}, \bibinfo {author} {\bibfnamefont {O.}~\bibnamefont {Dauchot}},\ and\ \bibinfo {author} {\bibfnamefont {D.}~\bibnamefont {Bartolo}},\ }\bibfield  {title} {\bibinfo {title} {Emergence of macroscopic directed motion in populations of motile colloids},\ }\href {https://doi.org/10.1038/nature12673} {\bibfield  {journal} {\bibinfo  {journal} {Nature}\ }\textbf {\bibinfo {volume} {503}},\ \bibinfo {pages} {95} (\bibinfo {year} {2013})}\BibitemShut {NoStop}%
\bibitem [{\citenamefont {Shani}\ \emph {et~al.}(2014)\citenamefont {Shani}, \citenamefont {Beatus}, \citenamefont {Bar-Ziv},\ and\ \citenamefont {Tlusty}}]{Shani2014}%
  \BibitemOpen
  \bibfield  {author} {\bibinfo {author} {\bibfnamefont {I.}~\bibnamefont {Shani}}, \bibinfo {author} {\bibfnamefont {T.}~\bibnamefont {Beatus}}, \bibinfo {author} {\bibfnamefont {R.~H.}\ \bibnamefont {Bar-Ziv}},\ and\ \bibinfo {author} {\bibfnamefont {T.}~\bibnamefont {Tlusty}},\ }\bibfield  {title} {\bibinfo {title} {Long-range orientational order in two-dimensional microfluidic dipoles},\ }\href@noop {} {\bibfield  {journal} {\bibinfo  {journal} {Nat. Phys.}\ }\textbf {\bibinfo {volume} {10}},\ \bibinfo {pages} {140} (\bibinfo {year} {2014})}\BibitemShut {NoStop}%
\bibitem [{\citenamefont {Nishiguchi}\ \emph {et~al.}(2017)\citenamefont {Nishiguchi}, \citenamefont {Nagai}, \citenamefont {Chat\'e},\ and\ \citenamefont {Sano}}]{Nishiguchi2017}%
  \BibitemOpen
  \bibfield  {author} {\bibinfo {author} {\bibfnamefont {D.}~\bibnamefont {Nishiguchi}}, \bibinfo {author} {\bibfnamefont {K.~H.}\ \bibnamefont {Nagai}}, \bibinfo {author} {\bibfnamefont {H.}~\bibnamefont {Chat\'e}},\ and\ \bibinfo {author} {\bibfnamefont {M.}~\bibnamefont {Sano}},\ }\bibfield  {title} {\bibinfo {title} {Long-range nematic order and anomalous fluctuations in suspensions of swimming filamentous bacteria},\ }\href {https://doi.org/10.1103/PhysRevE.95.020601} {\bibfield  {journal} {\bibinfo  {journal} {Phys. Rev. E}\ }\textbf {\bibinfo {volume} {95}},\ \bibinfo {pages} {020601} (\bibinfo {year} {2017})}\BibitemShut {NoStop}%
\bibitem [{\citenamefont {Baruah}\ \emph {et~al.}(2026)\citenamefont {Baruah}, \citenamefont {Padhan}, \citenamefont {Maji}, \citenamefont {Pandit},\ and\ \citenamefont {Sharma}}]{Baruah2026}%
  \BibitemOpen
  \bibfield  {author} {\bibinfo {author} {\bibfnamefont {P.~V.}\ \bibnamefont {Baruah}}, \bibinfo {author} {\bibfnamefont {N.~B.}\ \bibnamefont {Padhan}}, \bibinfo {author} {\bibfnamefont {B.}~\bibnamefont {Maji}}, \bibinfo {author} {\bibfnamefont {R.}~\bibnamefont {Pandit}},\ and\ \bibinfo {author} {\bibfnamefont {P.}~\bibnamefont {Sharma}},\ }\bibfield  {title} {\bibinfo {title} {First observation of turbulence-like state in dense algal suspensions},\ }\href {https://doi.org/10.1063/5.0314198} {\bibfield  {journal} {\bibinfo  {journal} {Phys. Fluids}\ }\textbf {\bibinfo {volume} {38}},\ \bibinfo {pages} {051904} (\bibinfo {year} {2026})}\BibitemShut {NoStop}%
\bibitem [{\citenamefont {Takaha}\ and\ \citenamefont {Nishiguchi}(2023)}]{Takaha2023}%
  \BibitemOpen
  \bibfield  {author} {\bibinfo {author} {\bibfnamefont {Y.}~\bibnamefont {Takaha}}\ and\ \bibinfo {author} {\bibfnamefont {D.}~\bibnamefont {Nishiguchi}},\ }\bibfield  {title} {\bibinfo {title} {Quasi-two-dimensional bacterial swimming around pillars: Enhanced trapping efficiency and curvature dependence},\ }\href {https://doi.org/10.1103/PhysRevE.107.014602} {\bibfield  {journal} {\bibinfo  {journal} {Phys. Rev. E}\ }\textbf {\bibinfo {volume} {107}},\ \bibinfo {pages} {014602} (\bibinfo {year} {2023})}\BibitemShut {NoStop}%
\bibitem [{\citenamefont {Spagnolie}\ \emph {et~al.}(2015)\citenamefont {Spagnolie}, \citenamefont {Moreno-Flores}, \citenamefont {Bartolo},\ and\ \citenamefont {Lauga}}]{Spagnolie2015}%
  \BibitemOpen
  \bibfield  {author} {\bibinfo {author} {\bibfnamefont {S.~E.}\ \bibnamefont {Spagnolie}}, \bibinfo {author} {\bibfnamefont {G.~R.}\ \bibnamefont {Moreno-Flores}}, \bibinfo {author} {\bibfnamefont {D.}~\bibnamefont {Bartolo}},\ and\ \bibinfo {author} {\bibfnamefont {E.}~\bibnamefont {Lauga}},\ }\bibfield  {title} {\bibinfo {title} {Geometric capture and escape of a microswimmer colliding with an obstacle},\ }\href {https://doi.org/10.1039/C4SM02785J} {\bibfield  {journal} {\bibinfo  {journal} {Soft Matter}\ }\textbf {\bibinfo {volume} {11}},\ \bibinfo {pages} {3396} (\bibinfo {year} {2015})}\BibitemShut {NoStop}%
\bibitem [{\citenamefont {Spagnolie}\ and\ \citenamefont {Lauga}(2012)}]{Spagnolie2012}%
  \BibitemOpen
  \bibfield  {author} {\bibinfo {author} {\bibfnamefont {S.~E.}\ \bibnamefont {Spagnolie}}\ and\ \bibinfo {author} {\bibfnamefont {E.}~\bibnamefont {Lauga}},\ }\bibfield  {title} {\bibinfo {title} {Hydrodynamics of self-propulsion near a boundary: predictions and accuracy of far-field approximations},\ }\href {https://doi.org/10.1017/jfm.2012.101} {\bibfield  {journal} {\bibinfo  {journal} {J. Fluid Mech.}\ }\textbf {\bibinfo {volume} {700}},\ \bibinfo {pages} {105–147} (\bibinfo {year} {2012})}\BibitemShut {NoStop}%
\bibitem [{\citenamefont {Brumley}\ \emph {et~al.}(2014)\citenamefont {Brumley}, \citenamefont {Wan}, \citenamefont {Polin},\ and\ \citenamefont {Goldstein}}]{Brumley2014}%
  \BibitemOpen
  \bibfield  {author} {\bibinfo {author} {\bibfnamefont {D.~R.}\ \bibnamefont {Brumley}}, \bibinfo {author} {\bibfnamefont {K.~Y.}\ \bibnamefont {Wan}}, \bibinfo {author} {\bibfnamefont {M.}~\bibnamefont {Polin}},\ and\ \bibinfo {author} {\bibfnamefont {R.~E.}\ \bibnamefont {Goldstein}},\ }\bibfield  {title} {\bibinfo {title} {Flagellar synchronization through direct hydrodynamic interactions},\ }\href {https://doi.org/10.7554/eLife.02750} {\bibfield  {journal} {\bibinfo  {journal} {eLife}\ }\textbf {\bibinfo {volume} {3}},\ \bibinfo {pages} {e02750} (\bibinfo {year} {2014})}\BibitemShut {NoStop}%
\bibitem [{\citenamefont {Pradipta}\ \emph {et~al.}(2026)\citenamefont {Pradipta}, \citenamefont {Lee}, \citenamefont {Tran}, \citenamefont {Welch}, \citenamefont {Sankar}, \citenamefont {Kim}, \citenamefont {Kumar}, \citenamefont {Yong}, \citenamefont {Hong}, \citenamefont {Lim},\ and\ \citenamefont {Cheng}}]{Pradipta2026}%
  \BibitemOpen
  \bibfield  {author} {\bibinfo {author} {\bibfnamefont {G.}~\bibnamefont {Pradipta}}, \bibinfo {author} {\bibfnamefont {W.}~\bibnamefont {Lee}}, \bibinfo {author} {\bibfnamefont {V.}~\bibnamefont {Tran}}, \bibinfo {author} {\bibfnamefont {K.}~\bibnamefont {Welch}}, \bibinfo {author} {\bibfnamefont {S.~K.}\ \bibnamefont {Sankar}}, \bibinfo {author} {\bibfnamefont {Y.}~\bibnamefont {Kim}}, \bibinfo {author} {\bibfnamefont {S.}~\bibnamefont {Kumar}}, \bibinfo {author} {\bibfnamefont {X.}~\bibnamefont {Yong}}, \bibinfo {author} {\bibfnamefont {J.}~\bibnamefont {Hong}}, \bibinfo {author} {\bibfnamefont {S.}~\bibnamefont {Lim}},\ and\ \bibinfo {author} {\bibfnamefont {X.}~\bibnamefont {Cheng}},\ }\bibfield  {title} {\bibinfo {title} {Seeing new depths: Three-dimensional flow of a free-swimming alga},\ }\href {https://doi.org/10.1103/2rr3-lbrn} {\bibfield  {journal} {\bibinfo  {journal} {Phys. Rev. X}\ }\textbf {\bibinfo {volume} {16}},\ \bibinfo {pages} {021019} (\bibinfo {year} {2026})}\BibitemShut {NoStop}%
\bibitem [{\citenamefont {Blake}(1971)}]{Blake1971}%
  \BibitemOpen
  \bibfield  {author} {\bibinfo {author} {\bibfnamefont {J.~R.}\ \bibnamefont {Blake}},\ }\bibfield  {title} {\bibinfo {title} {A note on the image system for a stokeslet in a no-slip boundary},\ }\href {https://doi.org/10.1017/S0305004100049902} {\bibfield  {journal} {\bibinfo  {journal} {Math. Proc. Camb. Philos. Soc.}\ }\textbf {\bibinfo {volume} {70}},\ \bibinfo {pages} {303–310} (\bibinfo {year} {1971})}\BibitemShut {NoStop}%
\bibitem [{\citenamefont {Mathijssen}\ \emph {et~al.}(2016)\citenamefont {Mathijssen}, \citenamefont {Doostmohammadi}, \citenamefont {Yeomans},\ and\ \citenamefont {Shendruk}}]{Mathijssen2016}%
  \BibitemOpen
  \bibfield  {author} {\bibinfo {author} {\bibfnamefont {A.~J. T.~M.}\ \bibnamefont {Mathijssen}}, \bibinfo {author} {\bibfnamefont {A.}~\bibnamefont {Doostmohammadi}}, \bibinfo {author} {\bibfnamefont {J.~M.}\ \bibnamefont {Yeomans}},\ and\ \bibinfo {author} {\bibfnamefont {T.~N.}\ \bibnamefont {Shendruk}},\ }\bibfield  {title} {\bibinfo {title} {Hydrodynamics of micro-swimmers in films},\ }\href {https://doi.org/10.1017/jfm.2016.479} {\bibfield  {journal} {\bibinfo  {journal} {J. Fluid Mech.}\ }\textbf {\bibinfo {volume} {806}},\ \bibinfo {pages} {35–70} (\bibinfo {year} {2016})}\BibitemShut {NoStop}%
\bibitem [{\citenamefont {Yoshinaga}\ and\ \citenamefont {Liverpool}(2018)}]{Yoshinaga2018}%
  \BibitemOpen
  \bibfield  {author} {\bibinfo {author} {\bibfnamefont {N.}~\bibnamefont {Yoshinaga}}\ and\ \bibinfo {author} {\bibfnamefont {T.~B.}\ \bibnamefont {Liverpool}},\ }\bibfield  {title} {\bibinfo {title} {From hydrodynamic lubrication to many-body interactions in dense suspensions of active swimmers},\ }\href@noop {} {\bibfield  {journal} {\bibinfo  {journal} {Eur. Phys. J. E}\ }\textbf {\bibinfo {volume} {41}},\ \bibinfo {pages} {76} (\bibinfo {year} {2018})}\BibitemShut {NoStop}%
\bibitem [{\citenamefont {Kim}\ \emph {et~al.}(2025)\citenamefont {Kim}, \citenamefont {Nagella}, \citenamefont {Choi},\ and\ \citenamefont {Takatori}}]{Kim2025}%
  \BibitemOpen
  \bibfield  {author} {\bibinfo {author} {\bibfnamefont {D.~Y.}\ \bibnamefont {Kim}}, \bibinfo {author} {\bibfnamefont {S.~G.}\ \bibnamefont {Nagella}}, \bibinfo {author} {\bibfnamefont {K.~H.}\ \bibnamefont {Choi}},\ and\ \bibinfo {author} {\bibfnamefont {S.~C.}\ \bibnamefont {Takatori}},\ }\bibfield  {title} {\bibinfo {title} {Direct experimental measurement of many-body hydrodynamic interactions with optical tweezers},\ }\href@noop {} {\bibfield  {journal} {\bibinfo  {journal} {Phys. Rev. Fluids}\ }\textbf {\bibinfo {volume} {10}},\ \bibinfo {pages} {064301} (\bibinfo {year} {2025})}\BibitemShut {NoStop}%
\bibitem [{\citenamefont {Kyoya}\ \emph {et~al.}(2015)\citenamefont {Kyoya}, \citenamefont {Matsunaga}, \citenamefont {Imai}, \citenamefont {Omori},\ and\ \citenamefont {Ishikawa}}]{Kyoya2015}%
  \BibitemOpen
  \bibfield  {author} {\bibinfo {author} {\bibfnamefont {K.}~\bibnamefont {Kyoya}}, \bibinfo {author} {\bibfnamefont {D.}~\bibnamefont {Matsunaga}}, \bibinfo {author} {\bibfnamefont {Y.}~\bibnamefont {Imai}}, \bibinfo {author} {\bibfnamefont {T.}~\bibnamefont {Omori}},\ and\ \bibinfo {author} {\bibfnamefont {T.}~\bibnamefont {Ishikawa}},\ }\bibfield  {title} {\bibinfo {title} {Shape matters: Near-field fluid mechanics dominate the collective motions of ellipsoidal squirmers},\ }\href {https://doi.org/10.1103/PhysRevE.92.063027} {\bibfield  {journal} {\bibinfo  {journal} {Phys. Rev. E}\ }\textbf {\bibinfo {volume} {92}},\ \bibinfo {pages} {063027} (\bibinfo {year} {2015})}\BibitemShut {NoStop}%
\bibitem [{SM_()}]{SM_ref}%
  \BibitemOpen
  \href@noop {} {}\bibinfo {note} {See Supplemental Material at [URL will be inserted by publisher], which includes detailed experimental and numerical procedures, additional experimental and numerical results, and supplemental videos from experiment and simulation.}\BibitemShut {Stop}%
\bibitem [{\citenamefont {Mondal}\ \emph {et~al.}(2021)\citenamefont {Mondal}, \citenamefont {Prabhune}, \citenamefont {Ramaswamy},\ and\ \citenamefont {Sharma}}]{Mondal2021}%
  \BibitemOpen
  \bibfield  {author} {\bibinfo {author} {\bibfnamefont {D.}~\bibnamefont {Mondal}}, \bibinfo {author} {\bibfnamefont {A.~G.}\ \bibnamefont {Prabhune}}, \bibinfo {author} {\bibfnamefont {S.}~\bibnamefont {Ramaswamy}},\ and\ \bibinfo {author} {\bibfnamefont {P.}~\bibnamefont {Sharma}},\ }\bibfield  {title} {\bibinfo {title} {Strong confinement of active microalgae leads to inversion of vortex flow and enhanced mixing},\ }\href {https://doi.org/10.7554/eLife.67663} {\bibfield  {journal} {\bibinfo  {journal} {eLife}\ }\textbf {\bibinfo {volume} {10}},\ \bibinfo {pages} {e67663} (\bibinfo {year} {2021})}\BibitemShut {NoStop}%
\bibitem [{foo()}]{footnote_flow_powerlaw}%
  \BibitemOpen
  \href@noop {} {}\bibinfo {note} {As an exception, the flow field in the front region does not display $\sim r^{-2}$ power-law decay, because it still could not be regarded as the far-field region due to the existence of two front flagella.}\BibitemShut {Stop}%
\bibitem [{\citenamefont {Leiderman}\ and\ \citenamefont {Olson}(2016)}]{Leiderman2016}%
  \BibitemOpen
  \bibfield  {author} {\bibinfo {author} {\bibfnamefont {K.}~\bibnamefont {Leiderman}}\ and\ \bibinfo {author} {\bibfnamefont {S.~D.}\ \bibnamefont {Olson}},\ }\bibfield  {title} {\bibinfo {title} {Swimming in a two-dimensional brinkman fluid: Computational modeling and regularized solutions},\ }\href {https://doi.org/10.1063/1.4941258} {\bibfield  {journal} {\bibinfo  {journal} {Phys. Fluids}\ }\textbf {\bibinfo {volume} {28}},\ \bibinfo {pages} {021902} (\bibinfo {year} {2016})}\BibitemShut {NoStop}%
\bibitem [{\citenamefont {Nagel}\ and\ \citenamefont {Gallaire}(2015)}]{Nagel2015}%
  \BibitemOpen
  \bibfield  {author} {\bibinfo {author} {\bibfnamefont {M.}~\bibnamefont {Nagel}}\ and\ \bibinfo {author} {\bibfnamefont {F.}~\bibnamefont {Gallaire}},\ }\bibfield  {title} {\bibinfo {title} {Boundary elements method for microfluidic two-phase flows in shallow channels},\ }\href {https://doi.org/https://doi.org/10.1016/j.compfluid.2014.10.016} {\bibfield  {journal} {\bibinfo  {journal} {Comput. Fluids}\ }\textbf {\bibinfo {volume} {107}},\ \bibinfo {pages} {272} (\bibinfo {year} {2015})}\BibitemShut {NoStop}%
\bibitem [{\citenamefont {Drescher}\ \emph {et~al.}(2010)\citenamefont {Drescher}, \citenamefont {Goldstein}, \citenamefont {Michel}, \citenamefont {Polin},\ and\ \citenamefont {Tuval}}]{Drescher2010}%
  \BibitemOpen
  \bibfield  {author} {\bibinfo {author} {\bibfnamefont {K.}~\bibnamefont {Drescher}}, \bibinfo {author} {\bibfnamefont {R.~E.}\ \bibnamefont {Goldstein}}, \bibinfo {author} {\bibfnamefont {N.}~\bibnamefont {Michel}}, \bibinfo {author} {\bibfnamefont {M.}~\bibnamefont {Polin}},\ and\ \bibinfo {author} {\bibfnamefont {I.}~\bibnamefont {Tuval}},\ }\bibfield  {title} {\bibinfo {title} {Direct measurement of the flow field around swimming microorganisms},\ }\href {https://doi.org/10.1103/PhysRevLett.105.168101} {\bibfield  {journal} {\bibinfo  {journal} {Phys. Rev. Lett.}\ }\textbf {\bibinfo {volume} {105}},\ \bibinfo {pages} {168101} (\bibinfo {year} {2010})}\BibitemShut {NoStop}%
\bibitem [{\citenamefont {Nakayama}\ and\ \citenamefont {Yamamoto}(2005)}]{Nakayama2005}%
  \BibitemOpen
  \bibfield  {author} {\bibinfo {author} {\bibfnamefont {Y.}~\bibnamefont {Nakayama}}\ and\ \bibinfo {author} {\bibfnamefont {R.}~\bibnamefont {Yamamoto}},\ }\bibfield  {title} {\bibinfo {title} {Simulation method to resolve hydrodynamic interactions in colloidal dispersions},\ }\href@noop {} {\bibfield  {journal} {\bibinfo  {journal} {Phys. Rev. E}\ }\textbf {\bibinfo {volume} {71}},\ \bibinfo {pages} {036707} (\bibinfo {year} {2005})}\BibitemShut {NoStop}%
\bibitem [{\citenamefont {Yamamoto}\ \emph {et~al.}(2021)\citenamefont {Yamamoto}, \citenamefont {Molina},\ and\ \citenamefont {Nakayama}}]{Yamamoto2021}%
  \BibitemOpen
  \bibfield  {author} {\bibinfo {author} {\bibfnamefont {R.}~\bibnamefont {Yamamoto}}, \bibinfo {author} {\bibfnamefont {J.~J.}\ \bibnamefont {Molina}},\ and\ \bibinfo {author} {\bibfnamefont {Y.}~\bibnamefont {Nakayama}},\ }\bibfield  {title} {\bibinfo {title} {Smoothed profile method for direct numerical simulations of hydrodynamically interacting particles},\ }\href@noop {} {\bibfield  {journal} {\bibinfo  {journal} {Soft Matter}\ }\textbf {\bibinfo {volume} {17}},\ \bibinfo {pages} {4226} (\bibinfo {year} {2021})}\BibitemShut {NoStop}%
\bibitem [{\citenamefont {Dauchot}\ \emph {et~al.}(2005)\citenamefont {Dauchot}, \citenamefont {Marty},\ and\ \citenamefont {Biroli}}]{Dauchot2005}%
  \BibitemOpen
  \bibfield  {author} {\bibinfo {author} {\bibfnamefont {O.}~\bibnamefont {Dauchot}}, \bibinfo {author} {\bibfnamefont {G.}~\bibnamefont {Marty}},\ and\ \bibinfo {author} {\bibfnamefont {G.}~\bibnamefont {Biroli}},\ }\bibfield  {title} {\bibinfo {title} {Dynamical heterogeneity close to the jamming transition in a sheared granular material},\ }\href {https://doi.org/10.1103/PhysRevLett.95.265701} {\bibfield  {journal} {\bibinfo  {journal} {Phys. Rev. Lett.}\ }\textbf {\bibinfo {volume} {95}},\ \bibinfo {pages} {265701} (\bibinfo {year} {2005})}\BibitemShut {NoStop}%
\bibitem [{\citenamefont {Caprini}\ \emph {et~al.}(2020)\citenamefont {Caprini}, \citenamefont {Marini Bettolo~Marconi},\ and\ \citenamefont {Puglisi}}]{Caprini2020}%
  \BibitemOpen
  \bibfield  {author} {\bibinfo {author} {\bibfnamefont {L.}~\bibnamefont {Caprini}}, \bibinfo {author} {\bibfnamefont {U.}~\bibnamefont {Marini Bettolo~Marconi}},\ and\ \bibinfo {author} {\bibfnamefont {A.}~\bibnamefont {Puglisi}},\ }\bibfield  {title} {\bibinfo {title} {Spontaneous velocity alignment in motility-induced phase separation},\ }\href {https://doi.org/10.1103/PhysRevLett.124.078001} {\bibfield  {journal} {\bibinfo  {journal} {Phys. Rev. Lett.}\ }\textbf {\bibinfo {volume} {124}},\ \bibinfo {pages} {078001} (\bibinfo {year} {2020})}\BibitemShut {NoStop}%
\end{thebibliography}%


\begin{thebibliography}{12}%
\makeatletter
\providecommand \@ifxundefined [1]{%
 \@ifx{#1\undefined}
}%
\providecommand \@ifnum [1]{%
 \ifnum #1\expandafter \@firstoftwo
 \else \expandafter \@secondoftwo
 \fi
}%
\providecommand \@ifx [1]{%
 \ifx #1\expandafter \@firstoftwo
 \else \expandafter \@secondoftwo
 \fi
}%
\providecommand \natexlab [1]{#1}%
\providecommand \enquote  [1]{``#1''}%
\providecommand \bibnamefont  [1]{#1}%
\providecommand \bibfnamefont [1]{#1}%
\providecommand \citenamefont [1]{#1}%
\providecommand \href@noop [0]{\@secondoftwo}%
\providecommand \href [0]{\begingroup \@sanitize@url \@href}%
\providecommand \@href[1]{\@@startlink{#1}\@@href}%
\providecommand \@@href[1]{\endgroup#1\@@endlink}%
\providecommand \@sanitize@url [0]{\catcode `\\12\catcode `\$12\catcode `\&12\catcode `\#12\catcode `\^12\catcode `\_12\catcode `\%12\relax}%
\providecommand \@@startlink[1]{}%
\providecommand \@@endlink[0]{}%
\providecommand \url  [0]{\begingroup\@sanitize@url \@url }%
\providecommand \@url [1]{\endgroup\@href {#1}{\urlprefix }}%
\providecommand \urlprefix  [0]{URL }%
\providecommand \Eprint [0]{\href }%
\providecommand \doibase [0]{https://doi.org/}%
\providecommand \selectlanguage [0]{\@gobble}%
\providecommand \bibinfo  [0]{\@secondoftwo}%
\providecommand \bibfield  [0]{\@secondoftwo}%
\providecommand \translation [1]{[#1]}%
\providecommand \BibitemOpen [0]{}%
\providecommand \bibitemStop [0]{}%
\providecommand \bibitemNoStop [0]{.\EOS\space}%
\providecommand \EOS [0]{\spacefactor3000\relax}%
\providecommand \BibitemShut  [1]{\csname bibitem#1\endcsname}%
\let\auto@bib@innerbib\@empty
\bibitem [{\citenamefont {Blair}\ and\ \citenamefont {Dufresne}(2008)}]{blair_matlab_tracking}%
  \BibitemOpen
  \bibfield  {author} {\bibinfo {author} {\bibfnamefont {D.}~\bibnamefont {Blair}}\ and\ \bibinfo {author} {\bibfnamefont {E.}~\bibnamefont {Dufresne}},\ }\href@noop {} {\bibinfo {title} {The {MATLAB} particle tracking code repository}} (\bibinfo {year} {2008}),\ \bibinfo {note} {available at https://site.physics.georgetown.edu/matlab/}\BibitemShut {NoStop}%
\bibitem [{\citenamefont {Thielicke}\ and\ \citenamefont {Stamhuis}(2014)}]{Thielicke2014}%
  \BibitemOpen
  \bibfield  {author} {\bibinfo {author} {\bibfnamefont {W.}~\bibnamefont {Thielicke}}\ and\ \bibinfo {author} {\bibfnamefont {E.~J.}\ \bibnamefont {Stamhuis}},\ }\bibfield  {title} {\bibinfo {title} {{PIVlab} -- towards user-friendly, affordable and accurate digital particle image velocimetry in {MATLAB}},\ }\href@noop {} {\bibfield  {journal} {\bibinfo  {journal} {J. Open Res. Softw.}\ }\textbf {\bibinfo {volume} {2}},\ \bibinfo {pages} {e30} (\bibinfo {year} {2014})}\BibitemShut {NoStop}%
\bibitem [{\citenamefont {Nagel}\ and\ \citenamefont {Gallaire}(2015)}]{Nagel2015}%
  \BibitemOpen
  \bibfield  {author} {\bibinfo {author} {\bibfnamefont {M.}~\bibnamefont {Nagel}}\ and\ \bibinfo {author} {\bibfnamefont {F.}~\bibnamefont {Gallaire}},\ }\bibfield  {title} {\bibinfo {title} {Boundary elements method for microfluidic two-phase flows in shallow channels},\ }\href {https://doi.org/https://doi.org/10.1016/j.compfluid.2014.10.016} {\bibfield  {journal} {\bibinfo  {journal} {Comput. Fluids}\ }\textbf {\bibinfo {volume} {107}},\ \bibinfo {pages} {272} (\bibinfo {year} {2015})}\BibitemShut {NoStop}%
\bibitem [{\citenamefont {Leiderman}\ and\ \citenamefont {Olson}(2016)}]{Leiderman2016}%
  \BibitemOpen
  \bibfield  {author} {\bibinfo {author} {\bibfnamefont {K.}~\bibnamefont {Leiderman}}\ and\ \bibinfo {author} {\bibfnamefont {S.~D.}\ \bibnamefont {Olson}},\ }\bibfield  {title} {\bibinfo {title} {Swimming in a two-dimensional brinkman fluid: Computational modeling and regularized solutions},\ }\href {https://doi.org/10.1063/1.4941258} {\bibfield  {journal} {\bibinfo  {journal} {Phys. Fluids}\ }\textbf {\bibinfo {volume} {28}},\ \bibinfo {pages} {021902} (\bibinfo {year} {2016})}\BibitemShut {NoStop}%
\bibitem [{\citenamefont {Liron}\ and\ \citenamefont {Mochon}(1976)}]{Liron1976}%
  \BibitemOpen
  \bibfield  {author} {\bibinfo {author} {\bibfnamefont {N.}~\bibnamefont {Liron}}\ and\ \bibinfo {author} {\bibfnamefont {S.}~\bibnamefont {Mochon}},\ }\bibfield  {title} {\bibinfo {title} {Stokes flow for a stokeslet between two parallel flat plates},\ }\href {https://doi.org/10.1007/BF01535565} {\bibfield  {journal} {\bibinfo  {journal} {J. Eng. Math.}\ }\textbf {\bibinfo {volume} {10}},\ \bibinfo {pages} {287} (\bibinfo {year} {1976})}\BibitemShut {NoStop}%
\bibitem [{\citenamefont {Batchelor}(1967)}]{Batchelor1967}%
  \BibitemOpen
  \bibfield  {author} {\bibinfo {author} {\bibfnamefont {G.~K.}\ \bibnamefont {Batchelor}},\ }\href@noop {} {\emph {\bibinfo {title} {An Introduction to Fluid Dynamics}}},\ Cambridge Mathematical Library\ (\bibinfo  {publisher} {Cambridge University Press},\ \bibinfo {year} {1967})\BibitemShut {NoStop}%
\bibitem [{\citenamefont {Jeanneret}\ \emph {et~al.}(2019)\citenamefont {Jeanneret}, \citenamefont {Pushkin},\ and\ \citenamefont {Polin}}]{Jeanneret2019}%
  \BibitemOpen
  \bibfield  {author} {\bibinfo {author} {\bibfnamefont {R.}~\bibnamefont {Jeanneret}}, \bibinfo {author} {\bibfnamefont {D.~O.}\ \bibnamefont {Pushkin}},\ and\ \bibinfo {author} {\bibfnamefont {M.}~\bibnamefont {Polin}},\ }\bibfield  {title} {\bibinfo {title} {Confinement enhances the diversity of microbial flow fields},\ }\href {https://doi.org/10.1103/PhysRevLett.123.248102} {\bibfield  {journal} {\bibinfo  {journal} {Phys. Rev. Lett.}\ }\textbf {\bibinfo {volume} {123}},\ \bibinfo {pages} {248102} (\bibinfo {year} {2019})}\BibitemShut {NoStop}%
\bibitem [{\citenamefont {Nakayama}\ and\ \citenamefont {Yamamoto}(2005)}]{Nakayama2005}%
  \BibitemOpen
  \bibfield  {author} {\bibinfo {author} {\bibfnamefont {Y.}~\bibnamefont {Nakayama}}\ and\ \bibinfo {author} {\bibfnamefont {R.}~\bibnamefont {Yamamoto}},\ }\bibfield  {title} {\bibinfo {title} {Simulation method to resolve hydrodynamic interactions in colloidal dispersions},\ }\href@noop {} {\bibfield  {journal} {\bibinfo  {journal} {Phys. Rev. E}\ }\textbf {\bibinfo {volume} {71}},\ \bibinfo {pages} {036707} (\bibinfo {year} {2005})}\BibitemShut {NoStop}%
\bibitem [{\citenamefont {Yamamoto}\ \emph {et~al.}(2021)\citenamefont {Yamamoto}, \citenamefont {Molina},\ and\ \citenamefont {Nakayama}}]{Yamamoto2021}%
  \BibitemOpen
  \bibfield  {author} {\bibinfo {author} {\bibfnamefont {R.}~\bibnamefont {Yamamoto}}, \bibinfo {author} {\bibfnamefont {J.~J.}\ \bibnamefont {Molina}},\ and\ \bibinfo {author} {\bibfnamefont {Y.}~\bibnamefont {Nakayama}},\ }\bibfield  {title} {\bibinfo {title} {Smoothed profile method for direct numerical simulations of hydrodynamically interacting particles},\ }\href@noop {} {\bibfield  {journal} {\bibinfo  {journal} {Soft Matter}\ }\textbf {\bibinfo {volume} {17}},\ \bibinfo {pages} {4226} (\bibinfo {year} {2021})}\BibitemShut {NoStop}%
\bibitem [{\citenamefont {Kim}\ and\ \citenamefont {Karrila}(1991)}]{kim&karrila}%
  \BibitemOpen
  \bibfield  {author} {\bibinfo {author} {\bibfnamefont {S.}~\bibnamefont {Kim}}\ and\ \bibinfo {author} {\bibfnamefont {S.}~\bibnamefont {Karrila}},\ }\href@noop {} {\emph {\bibinfo {title} {Microhydrodynamics: Principles and Selected Applications}}},\ Butterworth-Heinemann series in chemical engineering\ (\bibinfo  {publisher} {Elsevier Science \& Technology Books},\ \bibinfo {year} {1991})\BibitemShut {NoStop}%
\bibitem [{\citenamefont {Molina}\ \emph {et~al.}(2013)\citenamefont {Molina}, \citenamefont {Nakayama},\ and\ \citenamefont {Yamamoto}}]{Molina2013}%
  \BibitemOpen
  \bibfield  {author} {\bibinfo {author} {\bibfnamefont {J.~J.}\ \bibnamefont {Molina}}, \bibinfo {author} {\bibfnamefont {Y.}~\bibnamefont {Nakayama}},\ and\ \bibinfo {author} {\bibfnamefont {R.}~\bibnamefont {Yamamoto}},\ }\bibfield  {title} {\bibinfo {title} {Hydrodynamic interactions of self-propelled swimmers},\ }\href@noop {} {\bibfield  {journal} {\bibinfo  {journal} {Soft Matter}\ }\textbf {\bibinfo {volume} {9}},\ \bibinfo {pages} {4923} (\bibinfo {year} {2013})}\BibitemShut {NoStop}%
\bibitem [{\citenamefont {Delfau}\ \emph {et~al.}(2016)\citenamefont {Delfau}, \citenamefont {Molina},\ and\ \citenamefont {Sano}}]{Delfau2016}%
  \BibitemOpen
  \bibfield  {author} {\bibinfo {author} {\bibfnamefont {J.-B.}\ \bibnamefont {Delfau}}, \bibinfo {author} {\bibfnamefont {J.}~\bibnamefont {Molina}},\ and\ \bibinfo {author} {\bibfnamefont {M.}~\bibnamefont {Sano}},\ }\bibfield  {title} {\bibinfo {title} {Collective behavior of strongly confined suspensions of squirmers},\ }\href@noop {} {\bibfield  {journal} {\bibinfo  {journal} {Europhys. Lett.}\ }\textbf {\bibinfo {volume} {114}},\ \bibinfo {pages} {24001} (\bibinfo {year} {2016})}\BibitemShut {NoStop}%
\end{thebibliography}%
